\documentclass[12pt]{iopart}

\usepackage{color}
\usepackage{graphicx}
\usepackage{amssymb}

\def\blue{\textcolor{blue}}

\usepackage{hyperref} %%% Permet de faire les hyperlinks

\begin{document}
\title[
A coalescence  problem at criticality]
 {The critical behaviors and the scaling functions of a coalescence equation }

\author{
Xinxing Chen$^1$,
Victor Dagard$^2$, Bernard Derrida$^{2,3}$ and  Zhan Shi$^4$,}

\address{$^1$ School of Mathematical Sciences, Shanghai Jiaotong University, 200240 Shanghai, China, $\;$Partially supported by NSFC grants 11771286 and 11531001.$\;$}
\address{$^2$ Laboratoire de Physique, Ecole Normale Sup\'erieure, ENS, Universit\'e PSL, CNRS, Sorbonne Universit\'e, Universit\'e de Paris, F-75005 Paris, France}
\address{$^3$ Coll\`ege de France, Universit\'e PSL, 11 place Marcelin Berthelot, F-75231 Paris Cedex 05, France} 
\address{$^4$  Laboratoire de Probabilit\'es, Statistique  et  Mod\'elisation  (LPSM), Sorbonne Universit\'e, 4 place Jussieu, F-75252 Paris Cedex 05, France, }
\ead{chenxinx@sjtu.edu.cn,victor.dagard@phys.ens.fr,bernard.derrida@phys.ens.fr,\\ zhan.shi@upmc.fr }

\begin{abstract}
We show  that  a coalescence equation
exhibits a variety of critical behaviors, depending on the initial condition. This equation was introduced a few years ago to understand a toy model {studied by Derrida and Retaux to mimic} the depinning transition in presence of disorder. It was shown recently that  this toy model   exhibits
the same  critical behaviors as the  equation studied in the present work.
Here we find  several families of exact solutions of this coalescence  equation, in particular a family of scaling functions which  are closely related to the different possible critical behaviors.
These scaling functions lead to new conjectures, in particular on the shapes of the critical trees, that we have checked numerically.
\end{abstract}
%\keywords{???}
\ \\ \ \\
%   \centerline{\red{Dedicated to John Cardy on the occasion of his 70th birthday}}
\ \\ \ \\
\submitto{\jpa}
\maketitle
% \end{verbatim}
\normalsize

\section{Introduction}
The present work is totally devoted to the study of   the long time behavior  of a density   $f(x,t) \ge 0 $  on the positive real axis ($x \ge 0$)  which evolves according to
\begin{equation}
{{d f(x,t) \over dt} = {d f(x,t) \over dx} + {1\over 2} \int_0^x f(x-y,t) f(y,t) d y }
\label{recursion}
\end{equation}
(note that we do not require $f$ to be normalized).

\begin{figure}[h]
\centerline{\includegraphics[width=7.5cm]{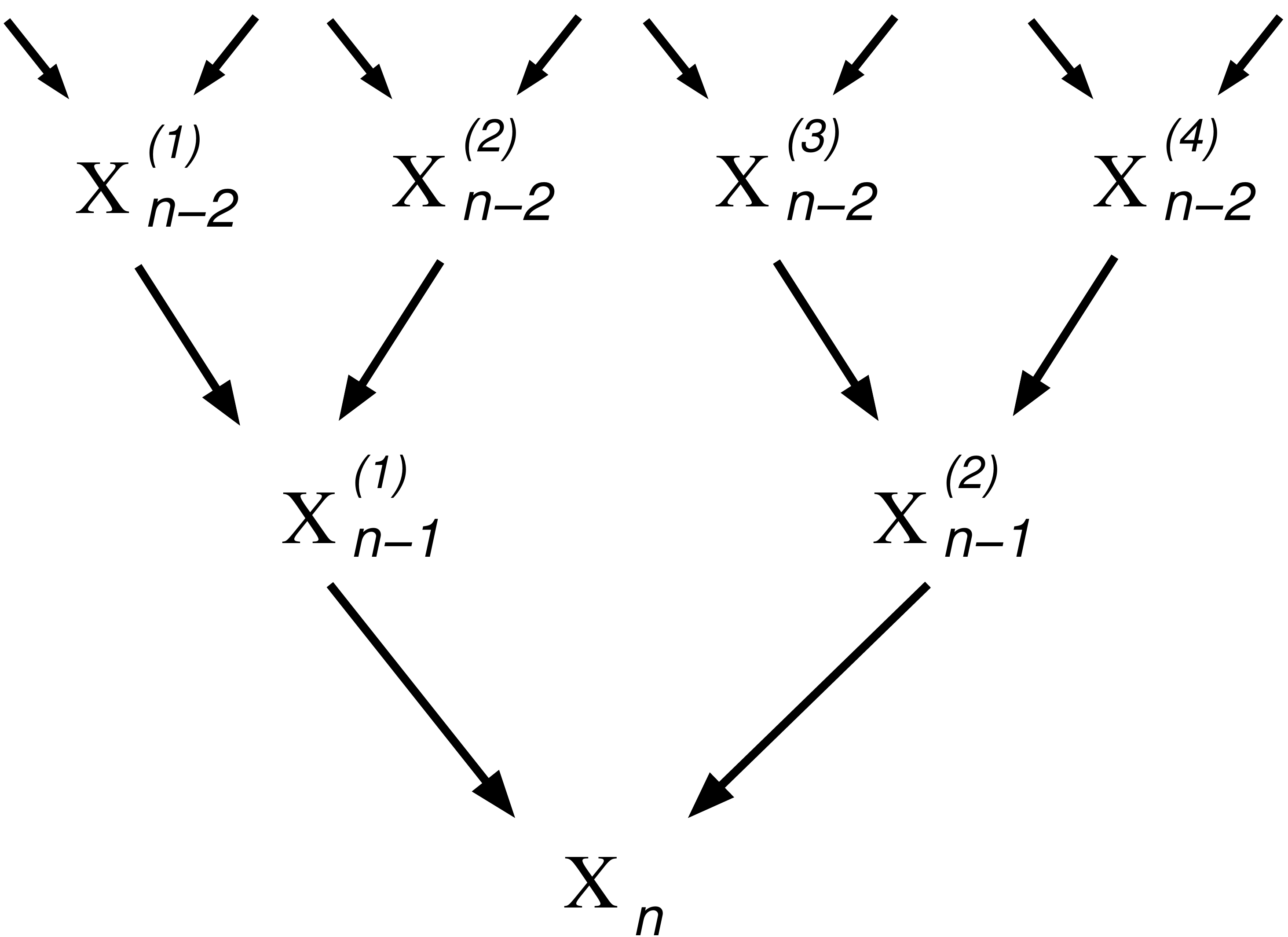}}
\caption{\small The random variable $X_n$ is   a determinisitic function (\ref{recur}) of  $2^n$ independent realizations of the random variable $X_0$ located at the top  of a binary tree.}
\label{tree1}
\end{figure}

This time evolution was introduced \cite{derrida-retaux} to analyze  
a simple renormalization problem  
\cite{collet-eckmann-glaser-martin2,
collet-eckmann-glaser-martin,derrida-retaux}
   which can be formulated as follows. Given a distribution $P(X_0)$ of a positive random variable $X_0$, what can be said  on the distribution of the random variable $X_n$  constructed through the following recursion 

\begin{equation}
X_n= \max\left[X_{n-1}^{(1) }+ X_{n-1}^{(2)} -1,0\right] 
\label{recur}
\end{equation}
where $X_{n-1}^{(1) } $ and  $X_{n-1}^{(2)} $ are two independent realizations of the variable $X_{n-1}$
(see Figure \ref{tree1}).

\ \\ \ \\
\ \\
\
This model was itself a simplified version of an old problem in the theory of disordered systems, the problem of depinning in presence of impurities \blue{\cite{luck,derrida-hakim-vannimenus,tang-chate,giacomin,giacomin-toninelli,dglt,monthus}}  and the relevant quantity (which plays the role of the free energy in the depinning problem) is the expectation of the free energy 
\begin{equation}
{\cal  F}_\infty = \lim_{n \to \infty} {1 \over 2^n} \sum_{k \ge 0}  k\  P(X_n=k) 
\label{free-energy}
\end{equation}
(the proof of the existence of this limit follows directly from the facts that $X_n \ge 0$ and that the sequence $\langle X_n \rangle \over 2^n$  is  decreasing). 
So the main question is to understand how the free-energy (\ref{free-energy}) depends on the initial distribution $P(X_0)$, in particular in the neighborhood of the  phase transition between a phase where ${\cal  F}_\infty=0$ and a phase where ${\cal  F}_\infty >0$.

\subsection{A short history of  recursion (\ref{recur})}
By far the easiest (non-trivial) case to consider is when the  initial distribution is concentrated on positive integer  values ($X_0 \ge 0$), in which case  it is easy to see from (\ref{recur}) that 
 the evolution of $Q_n(k)\equiv P(X_n=k)$ is given by 
\begin{equation}
Q_{n+1}(k) = 2 \Big(1-\sum_{k' \ge 1} Q_{n}(k') \Big) \ Q_{n} ( k+1) + \sum_{k'=1}^k Q_n(k') \ Q_n(k+1-k')
\label{recur-dist}
\end{equation}
and that 
the generating function of the distribution of $X_n$ 
$$H_n(z) = \sum_{k \ge 0} P(X_n=k) z^k$$
satisfies the following exact recursion
\begin{equation}
H_{n+1}(z)= {H_{n}(z)^2 - H_n(0)^2 \over z}  + H_n(0)^2 \ .
\label{rec1}
\end{equation}

This recursion was first studied long time ago by Collet, Eckmann, Glaser and Martin 
\cite{collet-eckmann-glaser-martin,collet-eckmann-glaser-martin2}
in the context of spins glasses.  Defining
\begin{equation}
\Delta  \equiv  2 H'_0(2)  - H_0(2)
\label{Delta-def}
\end{equation}
they were able  to determine the critical manifold
\begin{equation}
\Delta = 0 
\label{manifold}
\end{equation}
(so that $\Delta$ represents the distance to the critical manifold)
and to prove that 
\begin{equation}
\begin{array}{cccccc}
{\cal  F}_\infty
 &  = 0 & \ \ \ \ \ &{\rm   when } \ \ \ \ \ & \Delta  \le 0 &  \\
{\cal  F}_\infty
 &  > 0 & \ \ \ \ \ & {\rm   when } \ \ \ \ \ &  \Delta >0  & \  . 
\end{array}
\end{equation}
On the critical manifold (\ref{manifold})
they also conjectured that, for large $n $, 
\begin{equation}
1-P(X_n=0) = 1 - H_n(0) \simeq {4 \over n^2} \label{factor4}
\end{equation}
and that \cite{bmxyz_questions}
\begin{equation}
P(X_n=k | X_n \ne 0)   \to {1 \over 2^k}  \ \ \ \ \ {\rm  for} \ \ k \ge 1 \ \ . 
\label{crit2}
\end{equation}
\ \\ \

For example for a two-valued distribution of the form
\begin{equation}
P(X_0) = (1-p) \, \delta_{X_0} + p\,  \delta_{X_0-2} 
\label{two-delta}
\end{equation}
one has $H_0(z) =1-p + p z^2$
and the phase transition (\ref{Delta-def},\ref{manifold}) takes place at $p_c={1 \over 5} $ meaning for the free energy defined in (\ref{free-energy}) that  ${\cal  F}_\infty >0$ for ${p >  p_c}$ and ${\cal  F}_\infty=0$ for ${p \le p_c}$.

By relating  the problem (\ref{recur},\ref{free-energy}) to solutions of the continuous space-time equation (\ref{recursion}) (see also sections \ref{history} and \ref{neighborhood}), 
 the following critical behavior 
 was conjectured in  \cite{derrida-retaux,tang-chate}
\begin{equation}
{\cal  F}_\infty \sim  \exp \left[- \, {
% \text{Constant} \ 
\Delta^{-{1\over 2} +o(1)} }\right]  \ \ \ \ \ {\rm  as} \ \ \ \ \ \Delta \to 0^+
\label{critical-behavior}
\end{equation}
as the distance  $\Delta$ to the critical manifold 
vanishes.
\big(For  the two delta peak distribution (\ref{two-delta}), one has $\Delta=5\,  p - 1 \sim (p-p_c)$ and numerical evidence for the critical behavior (\ref{critical-behavior})
 was  shown in \cite{derrida-retaux}\big).

Trying to establish  the  conjecture (\ref{critical-behavior})  by a mathematical proof \cite{6authors} it appeared that     it was necessary to assume 
the following additional condition  on the critical  manifold  $\Delta = 0$ (see (\ref{manifold})) 
\begin{equation}
H_0'''(2) <  \infty
\label{condition}
\end{equation}

If this condition is not satisfied, in particular for distributions which have the following large $k$ decays
\begin{equation}
P(X_0= k) \simeq {{\rm  Constant} \over 2^k \ k^{\alpha}} \ \ \ \ \ \ \ {\rm  with} \ \  \ 2 < \alpha \le  4, 
\label{intro-alpha}
\end{equation}
then precise bounds were  obtained in \cite{6authors} which predict  that (\ref{critical-behavior}) becomes  
\begin{equation}
{\cal  F}_\infty \sim  \exp \left[-
%  \, {\rm  Constant}
  \ \Delta^{-{1 \over \alpha-2}+ o(1) } \right]  \ \ \ \ \ {\rm  as} \ \ \ \ \ \Delta \to 0^+  \ \ . 
\label{critical-behavior-alpha}
\end{equation}

\begin{figure}[h]
\centerline{\includegraphics[width=7.5cm]{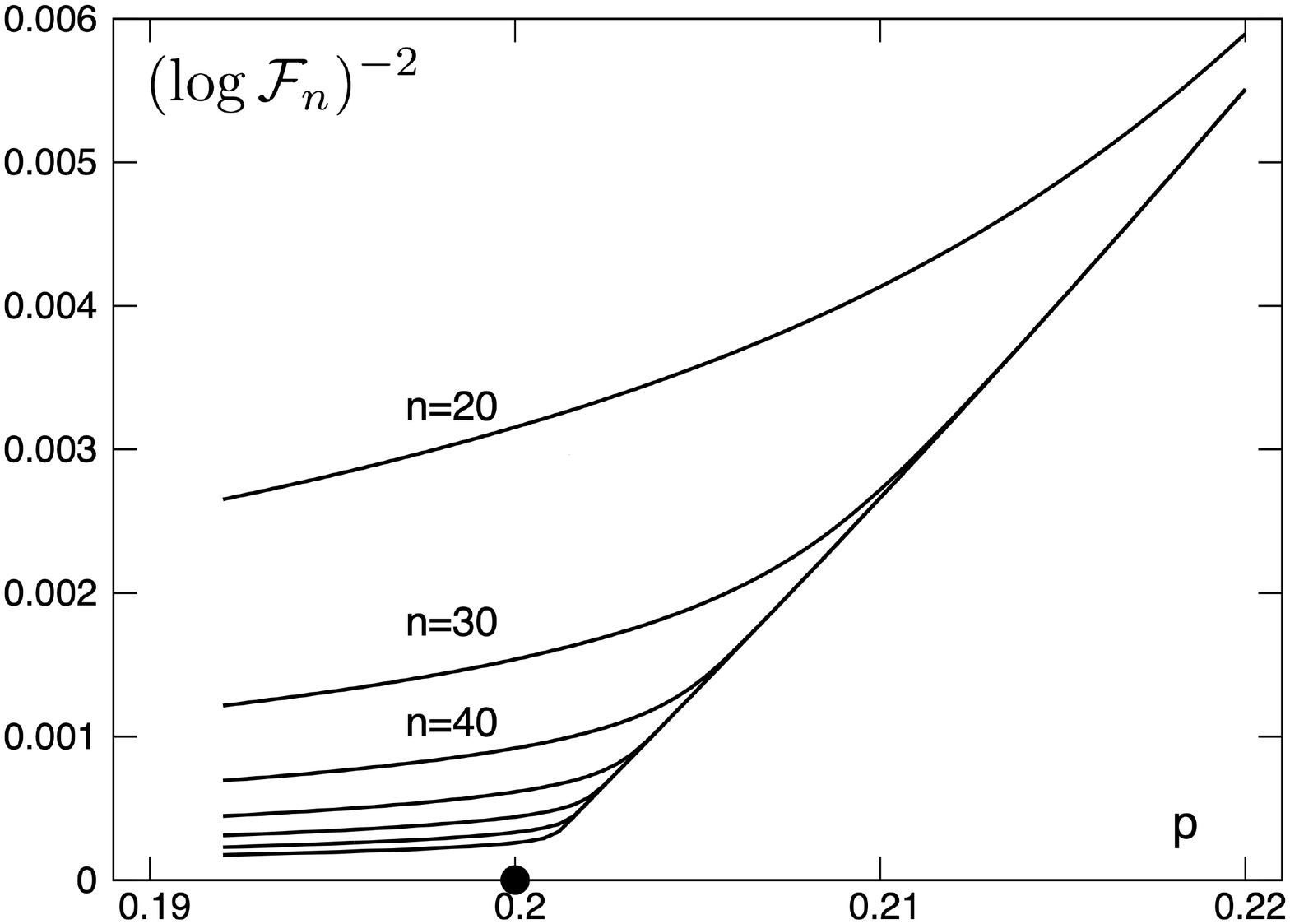}
 \ \ \ \includegraphics[width=7.5cm]{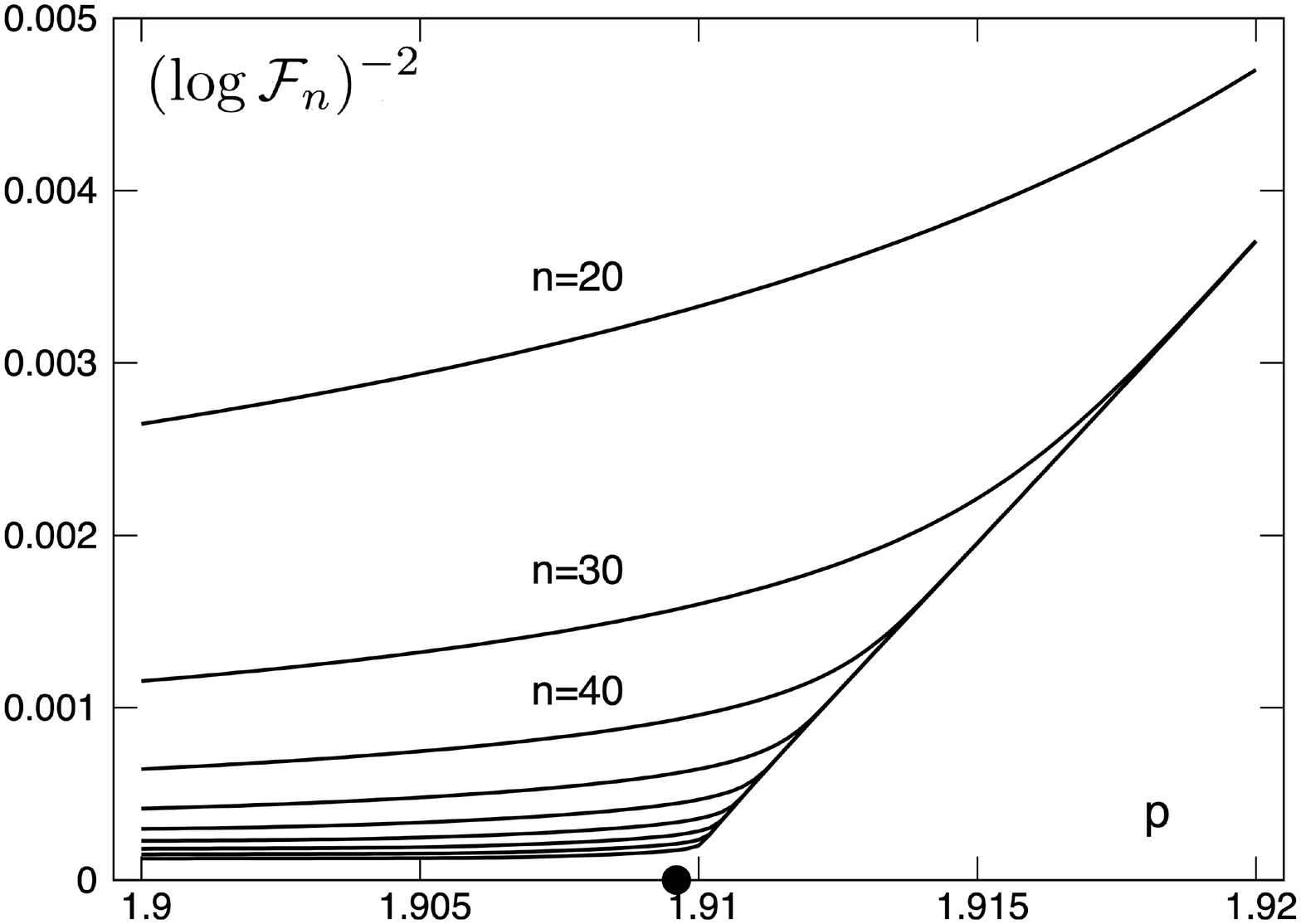}}
\caption{ \small According to (\ref{critical-behavior}), the plot $ (\log {\cal  F}_n)^{-2} $ versus $\Delta$ (here $\Delta$ is proportional to $p$)  should vanish linearly at the transition point $p_c$.  The two curves show this plot for the two distributions (\ref{two-delta}) and (\ref{alpha}) in the case $\alpha = 6$. (From (\ref{Delta-def},\ref{manifold}) the exact values of $p_c$ { shown by small black circles }can be determined: $p_c=.2$ and $p_c=  1.90956...$). As $n$ increases, the linearity of the plot looks better and bettter.}
\label{confirm1}
\end{figure}

\begin{figure}[h]
\centerline{\includegraphics[width=8cm]{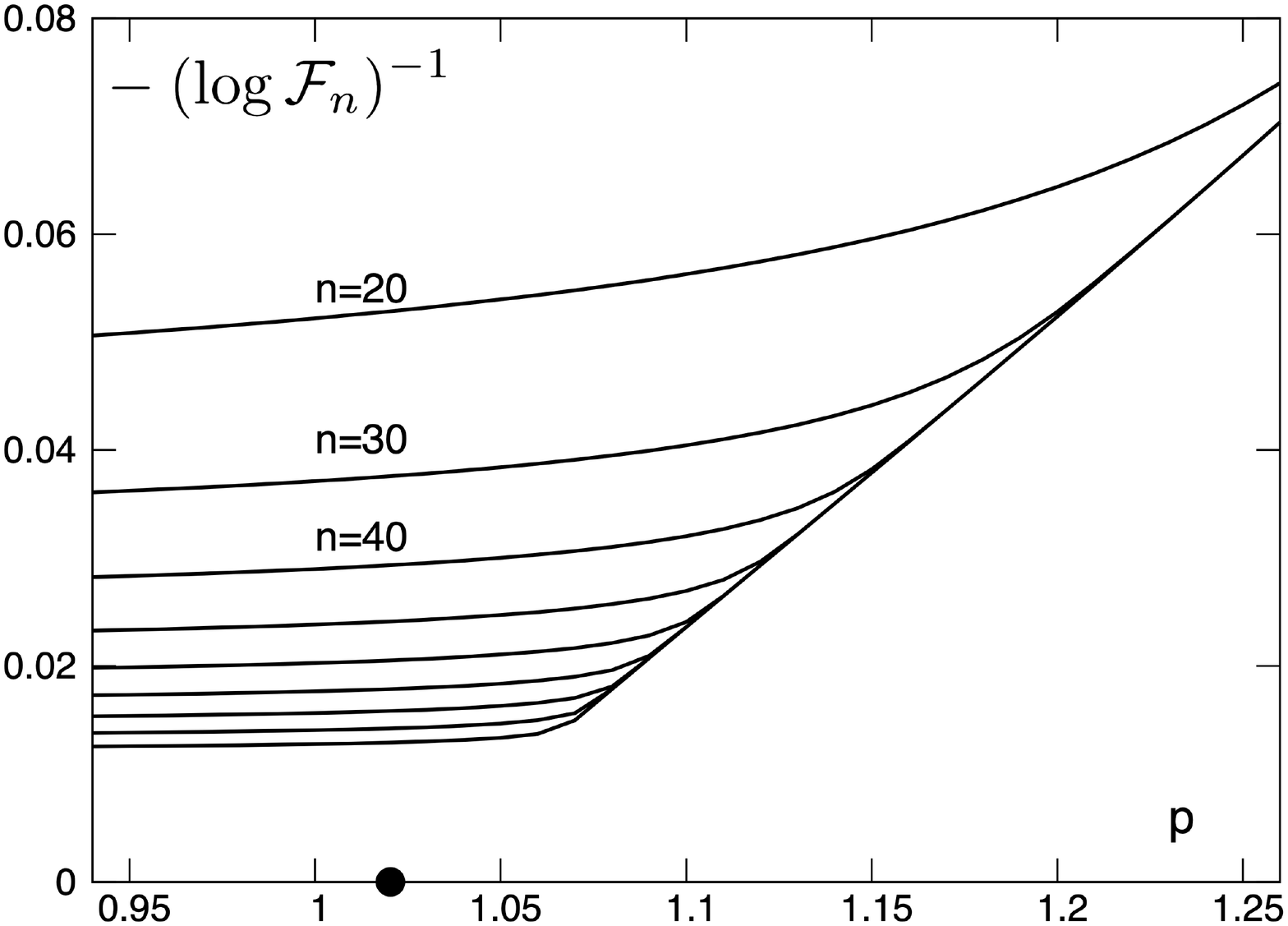}
 \ \ \ \includegraphics[width=8cm]{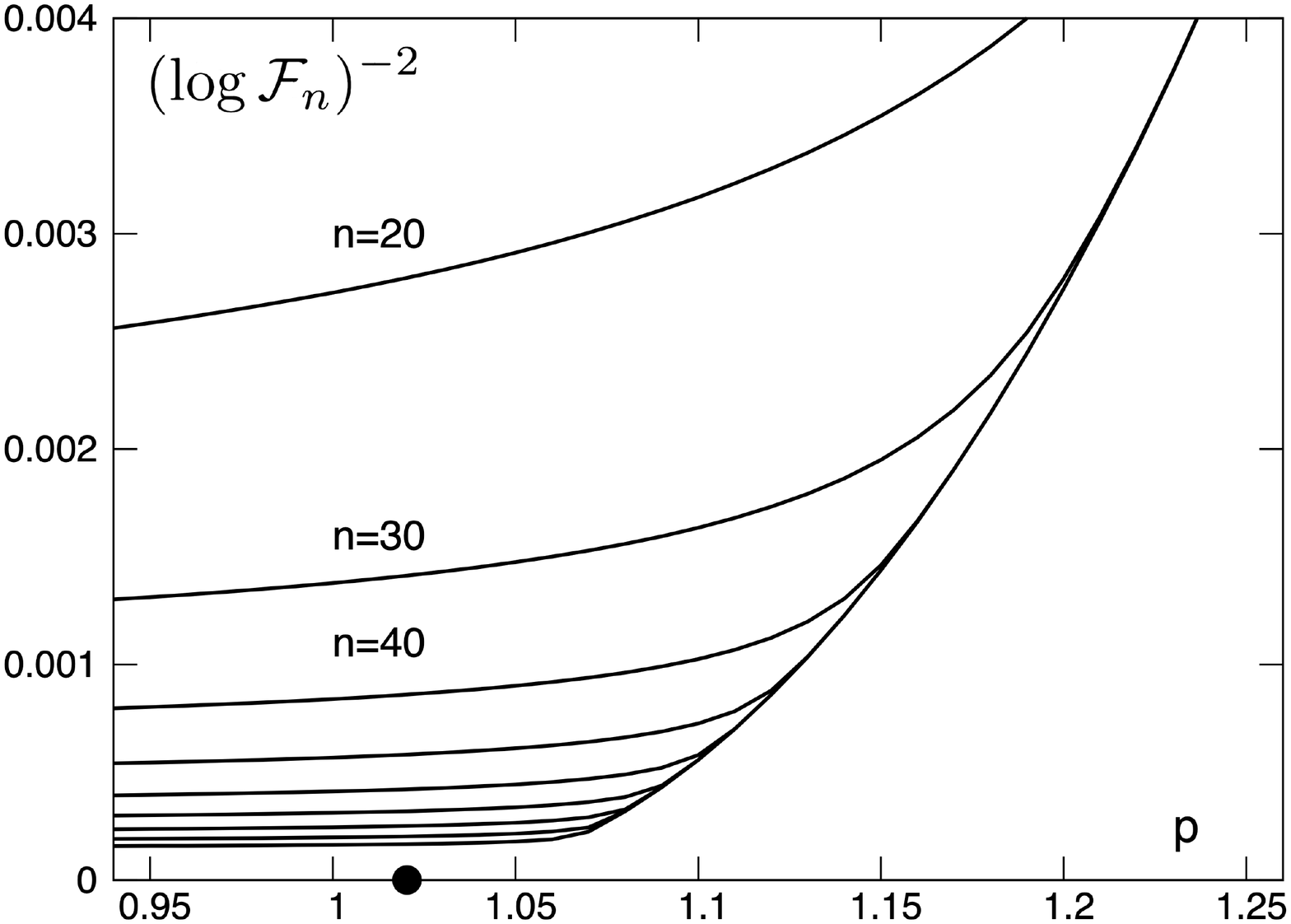}}
\caption{\small In contrast to the two  examples of Figure \ref{confirm1}, it is the plot of $ (-\log {\cal  F}_n)^{-1} $  which should be linear (see (\ref{critical-behavior-alpha})) for the distribution (\ref{alpha}) when $\alpha=3$. This behavior  is clearly seen in this case, in the vicinity of  the exact critical  point $p_c \simeq 1.02031...$. On the other hand, if one plots $ (\log {\cal  F}_n)^{-2} $ as in Figure \ref{confirm1},  the shape does not look linear  indicating that (\ref{critical-behavior}) is no longer valid. 
}
\label{confirm}
\end{figure}

The critical behaviors (\ref{critical-behavior},\ref{critical-behavior-alpha}) are confirmed in the
data of Figures \ref{confirm1} and \ref{confirm} where we show the results of exact numerical calculations of 
\begin{equation}
{\cal  F}_n = {1 \over 2^n} \sum_{k \ge 0} k \ P(X_n=k)
\label{Fn}
\end{equation}
as a function of $p$ for increasing values   of $n$ for the distribution (\ref{two-delta})
and for the distribution (\ref{alpha})
\begin{eqnarray}
P(X_0=k) &=  {p \over 2^k \ k^\alpha}   \ \ \ \ \ &  {\rm   for} \ k \ge 1 
\label{alpha}
\\ 
P(X_0=0) & = 1 -   \sum_{k\ge 1}  {p  \over 2^k \ k^\alpha}   \ \ \ \ \ &  {\rm   for} \ k  = 0  \ \ .
\nonumber
\end{eqnarray}

\ \\ {\it Remark:}
the above critical behaviors (\ref{critical-behavior},\ref{critical-behavior-alpha}) as well as the conjecture (\ref{factor4})  do remain valid \cite{6authors}  when, for any integer $m \ge 3$,  the recursion (\ref{recur}) is replaced  by
\begin{equation}
X_n= \max\left[X_{n-1}^{(1) }
+ X_{n-1}^{(2)} \cdots
+ X_{n-1}^{(m)} -1,0\right] 
\label{m-recur}
\end{equation}
provided that 
(\ref{manifold},\ref{factor4},\ref{condition},\ref{intro-alpha})
are modified
to become
$$
H_0(m)- m(m-1) H'_0(m) = 0 
\ \ \ \ \ ; \ \ \ \ 
 1 - H_n(0) \simeq {4 \over(m-1)^2 \  n^2} 
$$
and $$
H_0'''(m) <  \infty
\ \ \ \ \ ; \ \ \ \ 
P(X_0= k) \simeq {{\rm  Constant} \over m^k \ k^{\alpha}}  \ \ . 
$$

\subsection{Outline of the present paper}
One of the  goals of the present paper is to analyze  the case (\ref{intro-alpha}) from the perspective of 
the continuous equation (\ref{recursion}). We will see in Section \ref{particular solutions}  that, along the critical manifold ($\Delta=0$),  the distribution $P(X_n)$ takes a scaling form which depends on $\alpha$ for $2 < \alpha < 4$ and that this scaling form is described by     particular solutions of (\ref{recursion}). We will also see in Section \ref{neighborhood} how the $\alpha$-dependent critical behavior (\ref{critical-behavior-alpha}) emerges from the linearization of (\ref{recursion}) in the vicinity of the critical manifold. In section \ref{trees} we will discuss the shape of the tree connecting, at criticality and in the scaling regime,  some non zero value $ X_n$ to the initial values $X_0$.
But  we will start in section \ref{history}  by   recalling a few known facts about  the relation between the discrete problem (\ref{recur}) and  the continuous equation (\ref{recursion}).

\section{The relation between the continuous equation (\ref{recursion}) and the discrete problem (\ref{recur},\ref{free-energy})}
\label{history}

Based on the analysis  of numerical studies of the recursion (\ref{recur}) in \cite{derrida-retaux}
 it was noticed that, after a transient time and  in the neighborhood of the critical manifold ($\Delta \ll 1$),  the   distribution $P(X_n)$ evolves very slowly  and  that the data were consistent, for $k \ge 1$, with a scaling form
\begin{equation}
P(X_n=k) \equiv Q_n(k) \simeq {u^2 \over 2^k} \  f\left( u \, k ,  u \, n\right)
\label{sca1}
\end{equation} where $u$ is a small parameter.
 In the scaling regime, defining $x$ and $t$ by
\begin{equation}
 k={x \over u}  \ \ \ \ \ ;  \ \ \ \ n={t \over u}
\label{sca2}
\end{equation}
and inserting these forms into the recursion  (\ref{recur-dist}) 
for  $k \ge 1$
one obtains (\ref{recursion}) by keeping the leading order in $u$.

For distributions of the form (\ref{sca1}) one can  also  see (by keeping the leading order in $u$)  that the critical manifold (\ref{Delta-def},\ref{manifold})  becomes
\begin{equation}
\int_0^\infty  x f(x,0) dx = 1
\label{manifold1}
\end{equation}
and that (\ref{factor4},\ref{Delta-def},\ref{Fn}) 

\begin{equation}
1-P(X_n=0) \simeq u^2 f(0,{ u \, n })
\ \ \ \ \ ; \ \ \ \ \ 
\Delta= \int_0^\infty x f(x,0) dx -1  
\label{sca3}
\end{equation}
and
\begin{equation}
\label{sca4}
{\langle X_n \rangle \over 2^n}  \simeq 2^{1-{t \over u}} \  u^2\   f(0,t) \ \ . 
\end{equation}
(One can   check that (\ref{manifold1}) remains invariant under the evolution (\ref{recursion}).)

It was shown in \cite{derrida-retaux} that one particular solution of (\ref{recursion}) is
\begin{equation}
f(x,t) = {4 \kappa^2 \over \sin[\kappa (t+t_0)]^2}  \ \exp\left[ - {2 \kappa \,  x \over \tan[\kappa (t+t_0)]} \right] 
\label{expo1}
\end{equation}
where $\kappa$ and $t_0$ can be arbitrary ($\kappa$ could be real or  purely imaginary).
 For this distribution 
one has \begin{equation}
\Delta={1 \over \cos( \kappa  \, t_0)^2} -1
\label{expo2}
\end{equation}
 so that $\Delta >0$ corresponds to $\kappa$    real  (with $0< \kappa  \, t_0 < 2 \pi $), and $\Delta <0$ corresponds to $\kappa $ purely imaginary.
Along  the critical case $\Delta=0$ (given by $\kappa =0$)
\begin{equation}
f(x,t) = {4 \over  (t+t_0)^2}  \ \exp\left[ - {2 x \over t+t_0} \right] 
\label{expo3}
\end{equation}
and it is easy to see (\ref{sca2},\ref{sca3}) that (\ref{factor4})  is satisfied in the limit $n \to \infty$.

For $\Delta >0$ (i.e. for $\kappa $ real), it is clear that the solution (\ref{expo1}) diverges as  $t \to t_c$ 
where
\begin{equation}
t_c \equiv{\pi \over \kappa } - t_0 
\label{tc}
\end{equation} 
(because $\tan(\kappa(t+t_0)) \to 0^-$).

When $t$ approaches this limit, the scaling form  (\ref{sca1})  ceases to be valid (i.e. the system exits the scaling  regime and $ P(X_n)$  is no longer given by (\ref{sca1})). 
This can be seen in particular in the expression (\ref{sca3}) where  the divergence of $f(x,t)$ would make $P(X_n=0)$ become negative which is not possible. 
For $n \ge {t_c \over u}$, the  probability  $P(X_n=0)$ becomes small, so that one can forget the events where $X_n=0$  and one can replace the recursion (\ref{recur}) by simply $X_{n+1}=X_n^{(1)} + X_n^{(2)}-1$.
This leads to the following expression of ${\cal  F}_{\infty}$ defined by (\ref{free-energy})
when $t\to t_c$ (see (\ref{sca4}))
\begin{equation}
{\cal  F}_\infty\sim 
{\cal  F}_{n={t_c \over u}} \sim 
  2^{-{t_c\over u} }
\label{free-energy1}
\end{equation}
(one way to justify (\ref{free-energy1}) is to say that,   for $u$ small,  the scaling (\ref{sca4}) form remains valid as long as $t_c-t \geq u$).
As the limit $\Delta \to 0^+$ corresponds to the limit $\kappa  \to 0 $ \big(see (\ref{expo2})\big) which gives $\Delta \simeq {\kappa^2 t_0^2 \over2} $ one gets from (\ref{free-energy1}, \ref{tc})
\begin{equation}
{\cal  F}_\infty\sim   2^{-{\pi \over u \, \kappa } } \sim  2^{-{\pi \,  t_0 \over \sqrt{2} u} \, {1 \over \sqrt{\Delta}}}
\label{free-energy2}
\end{equation}
in agreement with (\ref{critical-behavior}).

\ \\ {\it Remark:} any initial condition $f(x,0)$ consisting of a single exponential can be written as (\ref{expo1}) by adjusting the parameters $\kappa$ and $t_0$.
For more general initial conditions $f(x,0)$, one expects  
that 
\begin{equation}
\begin{array}{ccccc} 
\nonumber
f(x,t) \to & \ 0 & \ \ \ \ \ \ &  {\rm  as} \ \ \  t \to \  \infty \ \ \  & {\rm  for} \ \ \ \Delta \le 0 \\
\label{f-tc}
f(x,t) \to & \ \infty  & \ \ \ \ \ &  {\rm  as}\ \ \  t \to \  t_c(\Delta)  \ \ \  & {\rm  for}\ \  \ \Delta > 0 
\end{array}
\end{equation}
depending on the sign of $\Delta$ defined in (\ref{sca3}).
% \begin{equation}
% \Delta =  \int_0^\infty  x f(x,0) dx \ \ - 1 \ \ . 
% \label{Delta-f}
% \end{equation}
Then, as in (\ref{free-energy1},\ref{free-energy2}), knowing how $t_c(\Delta)$ diverges as $\Delta \to 0$ allows one to predict the critical  behavior of ${\cal  F}_\infty$.

As for recursion (\ref{recur}) we will see that (\ref{free-energy2}) is expected to hold only when   condition (\ref{condition}) is fulfilled which, in the context of (\ref{recursion}), means that
\begin{equation}
\label{condition1}
\int_0^\infty x^3 \, f(x,0) dx < \infty \ \ . 
\end{equation}

\ \\ \ \\
{\it Remark:}
we will see in Section \ref{neighborhood}  that  both the critical behaviors (\ref{critical-behavior}) and (\ref{critical-behavior-alpha}) can also be understood by linearizing (\ref{recursion})  in the neighborhood of the scaling function (\ref{expo3}) as well as of the other scaling functions discussed in Section \ref{scaling functions}.

\ \\
{\it Remark:}
for distributions of the type (\ref{intro-alpha}),  when $\alpha < 2$
one has  $\Delta = \infty$ 
(see (\ref{Delta-def}))  
so that taking the limit $\Delta \to 0^+$ is meaningless.
One expects however \cite{yz_bnyz} in this case  that  for distributions of the form (\ref{alpha})
\begin{equation}
{\cal  F}_\infty \sim \exp\left[ - 
% \text{Constant}
\  p^{-{1 \over 2 - \alpha}+o(1)} \right] \label{alpha-less-than-2}
 \ \ \ \ \ \ {\rm as} \ \ p \to 0 \ \  . 
\end{equation}
From the view point of (\ref{recursion}) it is easy to check that, if $f(x,t)$  is the solution of (\ref{recursion}) for the initial  condition $f(x,0)= x^{-\alpha}$, then $f^*(x,t)= p^{2 \over 2 - \alpha} f(x \, p^{1 \over 2- \alpha} , t  \, p^{1 \over 2 - \alpha})$
is the solution of (\ref{recursion})  for the inititial condition $f^*(x,0)= p \, x^{-\alpha}$. Therefore if $f(x,t)$  blows up  at some critical time $t_c$, then $f^*(x,t)$ blows up at time $t_c^*= p^{-{1 \over 2 - \alpha}} \, t_c$ and repeating the reasoning going from (\ref{free-energy1}) to (\ref{free-energy2}) leads to (\ref{alpha-less-than-2}).

\ \\ \ \\
{\it Remark:}
motivated by the  discrete problem (\ref{recur}), Hu, Mallein and Pain proposed  in \cite{HMP}  a continuous space-time continuous version of the model  whose  evolution   differs   from (\ref{recursion})
\begin{equation}
{{d f(x,t) \over dt} = -f(x,t)+ {d f(x,t) \over dx} + {1\over 2} \int_0^x f(x-y,t) f(y,t) d y } \ \ . 
\label{HMP1}
\end{equation}
 For this problem too they were able to find  a family of exact solutions consisting of a single exponential   allowing  them to prove the critical behavior (\ref{critical-behavior}).

\section{ Families of exact solutions of (\ref{recursion})}
\label{particular solutions}
In \cite{derrida-retaux} the main predictions  (\ref{factor4},\ref{critical-behavior}) were based on the exact solution (\ref{expo1}) of (\ref{recursion}).
In this section we will exhibit several other families of solutions of (\ref{recursion}).
Our interest is limited to solutions of (\ref{recursion}) with non negative initial conditions ($f(x,0) \ge 0$) because $P(X_0=k)$ is a probability distribution (\ref{sca1}). Intuitively it is clear that, since (\ref{recursion})  was obtained through the scaling (\ref{sca1}), $f(x,t) $  should remain non-negative at $t>0$. A proof that evolving (\ref{recursion}) with a non-negative initial condition $f(x,0)$ leads to a non-negative solution $f(x,t)$ is given in  \ref{ap1}.

\subsection{The fixed points of (\ref{recursion})}

As shown in  \ref{ap2}, one can find a fixed point solution (i.e. a time independent solution) $f(x) $  of (\ref{recursion})  for any choice of $f(0)$.   All these solutions can be expressed in terms of a Bessel function whose sign varies   along the positive real axis. Therefore there is no way that such fixed point solutions can be reached or approached if one evolves (\ref{recursion}) starting with a non-negative initial condition $f(x,0)$. 
So we can forget these fixed point solutions.

\subsection{The scaling functions}
\label{scaling functions}
An important family of solutions of (\ref{recursion})  which will be  central in our understanding of (\ref{critical-behavior-alpha}) are scaling solutions of the form
\begin{equation}
f(x,t) = {1 \over (t+t_0)^2} F\left({x \over t+t_0} \right) \ \ . 
\label{F-scaling}
\end{equation}
They generalize (\ref{expo3}). Inserting (\ref{F-scaling}) into (\ref{recursion}) one gets immediately that  the scaling function $F$ should satisfy
\begin{equation}
 x F' + 2  F + F'
+  {1 \over 2}
\int_0^x F(y)   
 F(x-y)   \ d y 
=0 \ \ . 
\label{F-eq}
\end{equation}

\begin{figure}[h]
\centerline{\includegraphics[width=10.5cm]{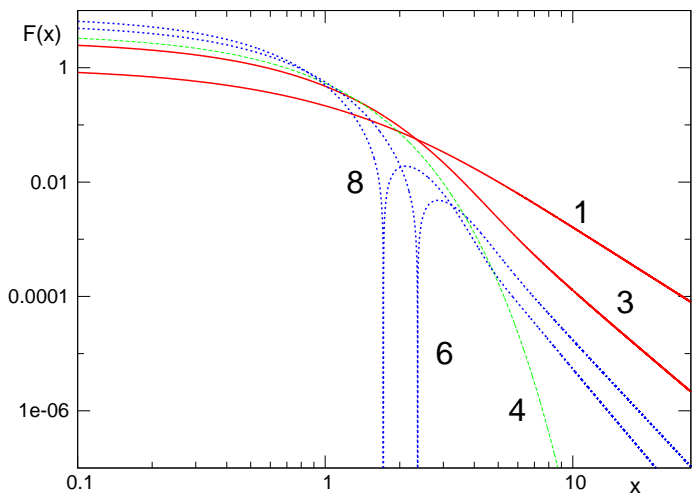}}
\caption{\small The  solution $F(x)$  
of (\ref{F-eq})
(or rather its absolute value $|F(x)|$ in the cases $F(0)=6$ or $8$) 
 for several choices ($F(0)= 1,3,4,6,8$) of $F(0)$. Except for $F(0)=4$ the large $x$ decay is a power law $F(x) \sim x^{-\alpha}$. We will see   (\ref{F-large-x}) that $\alpha= 1 + \sqrt{1 + 2 F(0)} $ . As here the figure shows a log-log plot of  $|F(x)|$, the zeroes of $F(x)$ appear as singularities  (where $\ln|F(x)| \to - \infty$) in the cases $F(0)=6$ or $8$.  }
\label{F-cur}
\end{figure}

For  an arbitrary value $F_0$ of $F(0)$ one can  solve (\ref{F-eq})  in powers of $x$ or in powers of  $F_0$:
\begin{eqnarray}
F(x) & = & F_0 - 2 F_0\  x + \left(3 F_0 - {F_0^2 \over 4} \right) x^2  +
 \left(-4 F_0 + {2F_0^2 \over 3} \right) x^3  +  O(x^4)
\nonumber
 \\ & = & {F_0  \over (1+ x)^2}  +F_0^2 \left( {x \over 2 (1+x)^2 (2+x)} - {\log (1+x) \over (2+x)^2}  \right)
+ O(F_0^3)  \ \ . 
\label{F0-expansion}
\end{eqnarray}
Apart from the special case $F_0=4$ for which $F(x)=4 e^{-2 x}$ (see (\ref{expo3}))
it is not clear whether  these series in powers of $x$ or in powers of $F_0$ converge, nor can one  tell from these expansions for which values of $F_0$, the scaling function $F$ remains non-negative.
One can however integrate numerically (\ref{F-eq}) as in Figure \ref{F-cur}. Except for $F_0=4$, one can observe a power law decay of $F(x)$  and that for $F_0 > 4$  the solution becomes negative.
\ \\ \ \\

In order to go further, it is easier to work with Laplace transforms. The Laplace transform $\widetilde{f}$  of $f$
\begin{equation}
\label{LP}
\widetilde{f}(p,t) = \int_0^\infty f(x,t)  \ e^{-p x} dx  
\end{equation}
 evolves   (see (\ref{recursion}))   according to
\begin{equation}
{d \widetilde{f}(p,t) \over dt} = - f(0,t) + p \widetilde{f}(p,t) +{1 \over 2} \widetilde{f}(p,t)^2  \ \ . 
\label{LP-evolution}
\end{equation}

For scaling solutions of the form (\ref{F-scaling}) 
 the  Laplace transform   $\widetilde{f}$  of $f$ 
 takes  also a  scaling form    
\begin{equation}
\widetilde{f}(p,t)  = {1 \over t+ t_0 } \widetilde{F} (p (t+t_0)) 
\label{LP-scaling}
\end{equation}
where $\widetilde{F}(q) $ satisfies
\begin{equation}
 \widetilde{F} +  q \widetilde{F} +{1 \over 2} \widetilde{F}^2 - F(0) - q \widetilde{F}' =0  \ \ . 
\label{eq2}
\end{equation} 
This is a non-linear equation! It turns out that it can be solved in terms of Bessel functions:
if one looks for a solution of the form
\begin{equation}
\widetilde{F}(q)= -1 - q -  q\  {y'({q\over 2} )\over y({q\over 2} )}
\label{F-Bessel}
\end{equation}
one gets from (\ref{eq2}) 
that $y(q)$ should satisfy
\begin{equation}
q^2 y'' + q y' - \left( q^2 + \beta^2 \right) y=0
\ \ \ \ \ {\rm  where} \ \ 
\beta^2 = {1 \over 4} + {F(0)\over 2}  \ \ . 
\label{Bessel equation}
\end{equation}
As $y$ is solution of a second order equation, it depends a priori on two arbitrary constants.
For example for large  $q$  it  depends on the two constants $B$ and $B'$:
\begin{eqnarray}
y= & B \ { e^{-q} \over \sqrt{q}} \left(1 + {4 \beta^2 -1\over 8 \ q }  + 
{(4 \beta^2-1)( 4 \beta^2-9) \over 128 \ q^2} 
 + \cdots \right) 
\label{y-large-q}
\\ 
+& B' \ { e^{q} \over \sqrt{q}} \left(1 - {4 \beta^2 -1\over 8 \ q }  + 
{(4 \beta^2-1)( 4 \beta^2-9) \over 128 \ q^2}  \ \ . 
 + \cdots \right)  \ \ . 
\nonumber
\end{eqnarray}

If $B' \ne 0$ then
 $\widetilde{F}(q) \simeq -2 q $ for large $q$ which cannot be as 
 $\widetilde{F}(q) $  is the Laplace transform of a non-negative   function.
Therefore $B'=0$. Moreover only the logarithmic derivative of $y$ is needed (see (\ref{F-Bessel}))  so that the choice of the constant $B$ does not matter. Therefore one can choose for $y$ the modified Bessel function $K_\beta$:
\begin{equation}
y=   K_\beta(q) \equiv   \int_0^\infty dt \, \cosh(\beta t) \ \exp[-q \cosh(t) ] \ \ .
\label{Kbeta}
\end{equation}
From (\ref{y-large-q})  (with $B'=0$) one gets for large $q$
$$\widetilde{F}(q) = {4 \beta^2-1  \over 2 q} -{4 \beta^2-1 \over q^2}  - {(4 \beta^2-1) (4 \beta^2-25)\over 8 q^3}  + \cdots
    $$
This coincides, as it should,  with the  large $q$ expansion which can be obtained directly from (\ref{eq2})
$$\widetilde{F}(q) = {F(0) \over q} -{2 F(0)\over q^2}  + {F(0) (12- F(0))\over 2 q^3}  + \cdots $$
when $F(0)$ and $\beta$ are related as in (\ref{Bessel equation}).
\ \\ \ \\
For small $q$, one can also show from (\ref{Kbeta}) that
\small{
\begin{eqnarray*}
y(q) \simeq  & \ 
  2^{\beta -1} \   q^{-\beta    }
 \left(\Gamma(\beta)  - q^2 {\Gamma(\beta-1) \over 4 }+ q^4 {\Gamma(\beta-2)  \over 32 }
 + \cdots\right)
\\ & +
  2^{-\beta -1} \   q^{\beta    }
 \left(\Gamma(-\beta)  - q^2 {\Gamma(-\beta-1) \over 4 }+ q^4 {\Gamma(-\beta-2)  \over 32 }
 + \cdots\right)
\end{eqnarray*}
which gives using (\ref{F-Bessel})
\begin{eqnarray}
\widetilde{F}(q) = & \left(  (2 \beta-1)  -   q + { q^2 \over 4 (\beta-1)} - {   q^4\over 64  (\beta-1)^2 (\beta-2) } + \cdots 
\right) 
\label{F-laplace}
 \\ & + \ c(\beta) \,   q^{2 \beta } \left( 1 + { q^2 \over 8 (\beta-1)} 
%+ {q^4 (2 \beta-5) \over 256 (\beta-1)^2 (\beta-2)} 
 +\cdots \right)  
% \\ 
 \ + \ c(\beta)^2 \,   q^{4 \beta } \left( {1 \over 4 \beta}  
% + { q^2 (2 \beta +1)\over 32  (\beta^2-1) \beta } 
+\cdots \right)  
+ \cdots 
\nonumber
\end{eqnarray}
where 
\begin{equation}
c(\beta) =  2^{2-4 \beta}  {\Gamma(1-\beta) \over \Gamma(\beta)}  \ \ .
\label{c-beta}
\end{equation}

The non-analytic term in the small $q$ expansion determines the large $x$ decay of the scaling function $F(x)$
\begin{equation}
\hspace{-2cm} 
F(x) \simeq {c(\beta) \over \Gamma(-2 \beta) \ x^{1+2 \beta}} = {2^{4- 2 \alpha} \Gamma({3-\alpha \over 2} ) \over \Gamma(1-\alpha) \ \Gamma({\alpha-1 \over 2})} {1 \over x^\alpha}  \ \ \ \ \ {\rm  where} \ \ \alpha=1+2 \beta = 1 + \sqrt{1+2F(0)} \ \ \ 
 \label{F-large-x}
\end{equation}
(see (\ref{Bessel equation})). So varying $F(0)$, i.e. varying $\beta$, changes the power-law decay 
of the scaling function $F$.
\ \\ \ \\
{\it Remark:} in (\ref{F-laplace}) one should reorder  the terms in the small $q$ expansion depending on the value of $\beta$. For example for ${1 \over 2} < \beta < 1$
$$\widetilde{F}(q)
 =( 2 \beta -1) - q + c(\beta) \, q^{2 \beta} + O(q^2)$$
whereas for $1 < \beta < {3 \over 2} $
$$\widetilde{F}(q) =   (2 \beta-1)  -   q + { q^2 \over 4 (\beta-1)} 
  + \ c(\beta) \   q^{2 \beta }  + o(q^3) \ \ .$$
\ \\ \ \\
{\it Remark:} we did not treat the cases where $\beta$ is an integer or half an integer.  They  could be analyzed  as limiting cases of (\ref{F-laplace}). For half integer values of $\beta$ there are only a finite number of terms in the sums (\ref{y-large-q}) and  $\widetilde{F}$ is a rational function. For example 

\begin{eqnarray}
\nonumber
\widetilde{F} (q) & =  {4 \over q+2}  \ \ \ \ \ &  {\rm  for} \ \ \beta={3 \over 2}   \\
\label{particular}
\widetilde{F} (q) & =  {12(q+4 )  \over q^2+6 q +12}  \ \ \ \ \  & {\rm  for} \ \ \beta={5 \over 2}   \\
\widetilde{F} (q) & =  {24(q^2 +10 q+30 )  \over q^3+12 q^2+60 q +120}  \ \ \ \ \  & {\rm  for} \ \ \beta={7 \over 2}   \ \ .
\nonumber
\end{eqnarray}

Except for the case $\beta={3 \over 2}$  which corresponds to the scaling function (\ref{expo3}), these solutions do not remain non-negative   along the whole  positive real axis. For example for $\beta={5 \over 2} $ one gets
\begin{equation}
F(x) = 12 e^{-3 x} \cos( \sqrt{3} x) + 4 \sqrt{3}  e^{-3 x} \sin(\sqrt{3} x)
\label{particular1}
\end{equation}
which obviously does not remain positive along the whole real axis. So as for the fixed points of (\ref{recursion}) one can forget these solutions.

 The particular solutions (\ref{particular})
correspond to the following expressions of $y(q)$

\begin{eqnarray*}
y(q) & = {q+1 \over q^{3\over 2}} \ e^{-q}  \ \ \ \ \ &  {\rm  for} \ \ \beta={3 \over 2}   \\
 & =  { q^2+3 q +3 \over q^{5\over 2}} \  e^{-q}  \ \ \ \ \  & {\rm  for} \ \ \beta={5 \over 2}  \\
 & =  { q^3+6 q^2+15 q +15 \over q^{7 \over 2}} \  e^{-q} \ \ \ \ \  & {\rm  for} \ \ \beta={7 \over 2}   \ \ . 
\end{eqnarray*}
\ \\ \ \\
{\it Remark:}  we believe, but  did not succeed to prove from (\ref{F-Bessel}),  that $\widetilde{F}(q)$ is the Laplace transform of a non-negative   function $F(x)$ when  $ 2 < \alpha \le 4$. One  can however see that the tail  (\ref{F-large-x}) is negative for 
$0\le  \alpha < 2$, 
$4 < \alpha < 6$, 
$8 < \alpha < 10$, 
  etc .  Moreover for $\alpha >4$, that is for $\beta > 3/2$, the
expansion (\ref{F-laplace}) gives  $
\widetilde{F}(q) =  2 \beta-1  -   q + { q^2 \over 4 (\beta-1)} + o(q^3)$
which implies that $\int_0^\infty x^3 F(x) dx =0$. Therefore  the scaling function cannot be non-negative   on the entire positive real axis for $\alpha > 4$. (In Figure \ref{F-cur} it was already clear that the scaling function $F(x)$ has at least one zero for $\alpha >4$.)
\ \\ \ \\
{\it Remark:}  it is easy to see  that  for $2 < \alpha \le 4$ (remember (\ref{F-large-x}) that $ \alpha=1+ 2 \beta = 1 + \sqrt{1+ 2 F(0)}$), the large $t$ decay of the scaling solutions (\ref{F-scaling}) is
$$f(0,t) \simeq  {F(0) \over t^2} = { \alpha (\alpha-2) \over 2 \, t^2}$$
If one comes back (see (\ref{sca1})) to $P(X_n)$   this means that
\begin{equation}
1- P(X_n=0) =  \sum_{k \ge 1} P(X_n=k) \simeq  u^2 f(0,t)  \simeq {\alpha (\alpha-2) \over 2 \,  n^2}
\label{factor-alpha}
\end{equation}
to be compared with (\ref{factor4}) when condition (\ref{condition}) is fulfilled.

Figure \ref{P(Xn=0)} shows the product $n^2 \Big(1 - P(X_n=0)\Big)$ versus $1/n$ 
\Big(with $P(X_n=k)$ obtained by iterating (\ref{recur})\Big) for the three initial conditions already considered in Figures \ref{confirm1}  and \ref{confirm}. For the two delta-peak distribution and for the distribution (\ref{alpha}) with $\alpha =6$ the data confirm (\ref{factor4}), indicating that  the large $n$ asymptotics of $P(X_n)$  is described by (\ref{expo3}).  When condition (\ref{condition}) is not satisfied,  here in the case of (\ref{alpha}) with $\alpha=3$, the data are consistent with (\ref{factor-alpha})  suggesting that the asymptotics follow the $\alpha$-dependent scaling function $F(x)$.

\begin{figure}[h]
\centerline{\includegraphics[width=10cm]{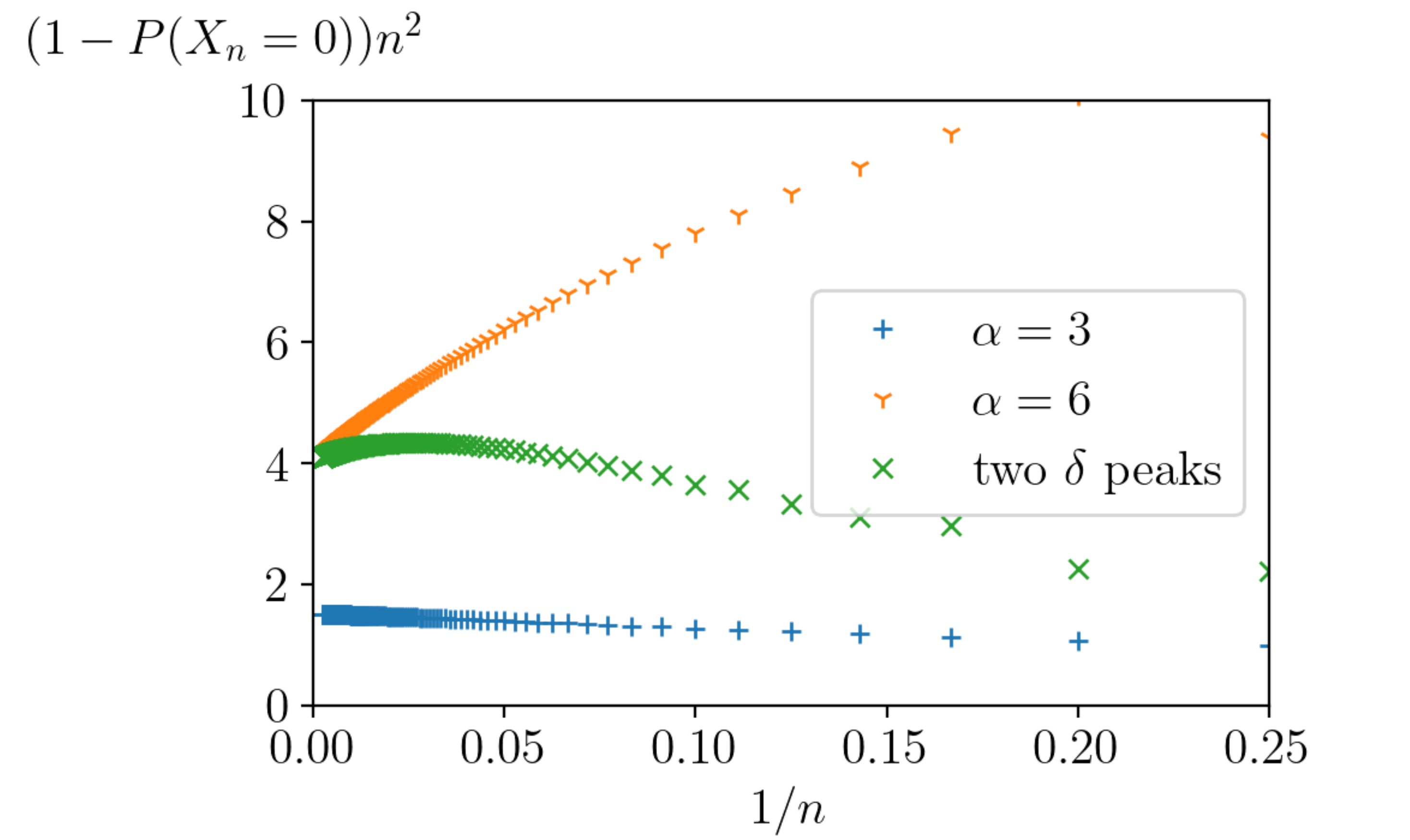}
}
\caption{\small For the two delta peak distribution (\ref{two-delta}) (at criticality i.e. for $p={1\over 5}$)  as well for the distribution (\ref{alpha}) when $\alpha=6$ (at criticality i.e. for $p=1.90956...$),  which both satisfy condition (\ref{condition}), the large $n$ asymptotics agrees with the prediction (\ref{factor4}) that $n^2 (1 - P(X_n=0)) \to 4 $. On the other hand for (\ref{alpha})  with $\alpha =3$ (at criticality i.e. for $p=1.02031...$) which does not satisfy condition (\ref{condition}), one observes  the asymptotics (\ref{factor-alpha}), i.e.  $n^2 (1-P(X_n=0)) \to \alpha(\alpha-2)/2$.
%\red{METTRE LE P EN ITALIQUES ET EVENTUELLEMENT CHANGER LA TAILLE DES FONTS POUR QU'ELLES RESSEMBLENT AUX AUTRES FIGURES}
}
\label{P(Xn=0)}
\end{figure}

\begin{figure}[h!]
\includegraphics[width=8cm]{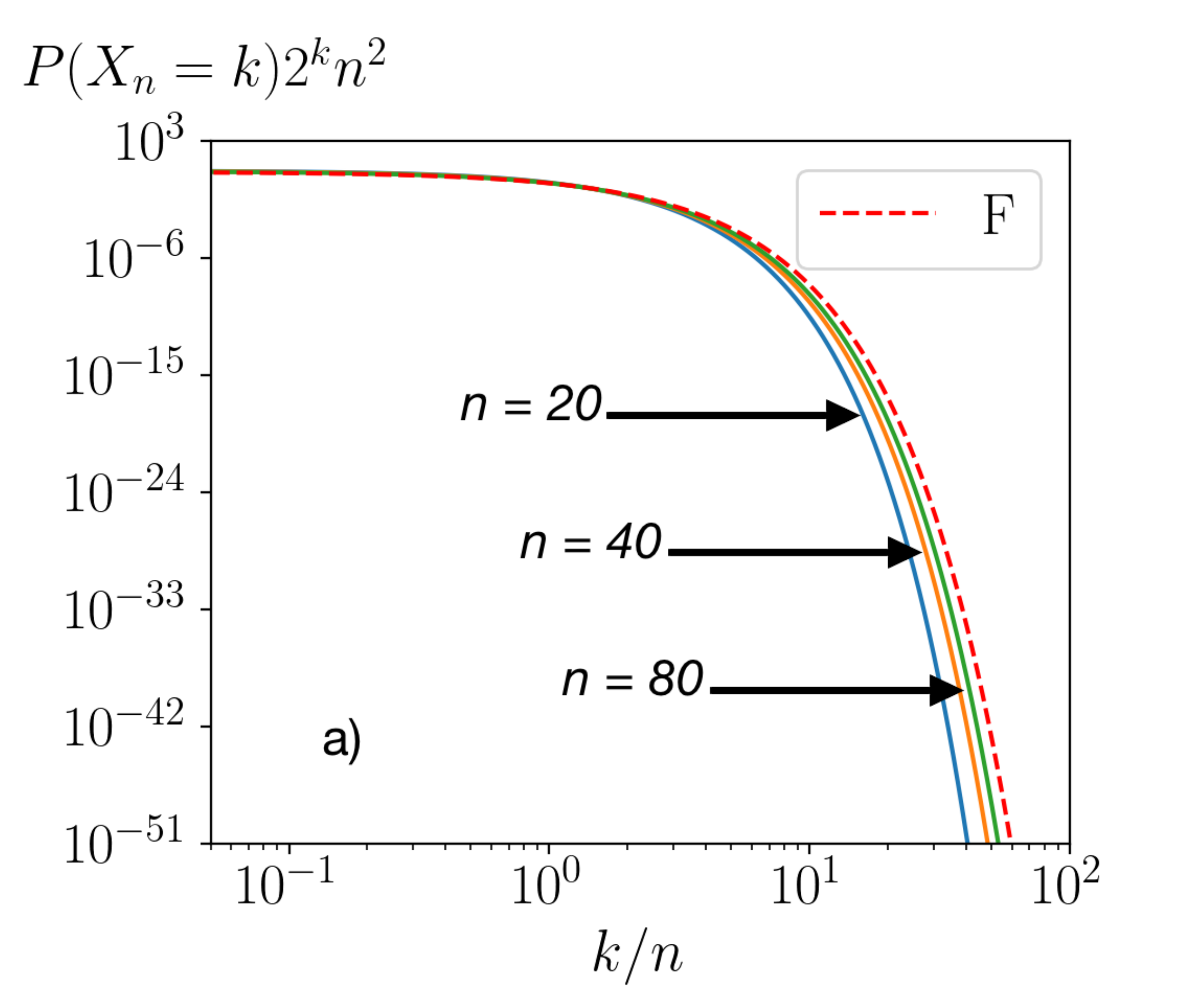} 
\ \ \ 
\includegraphics[width=8cm]{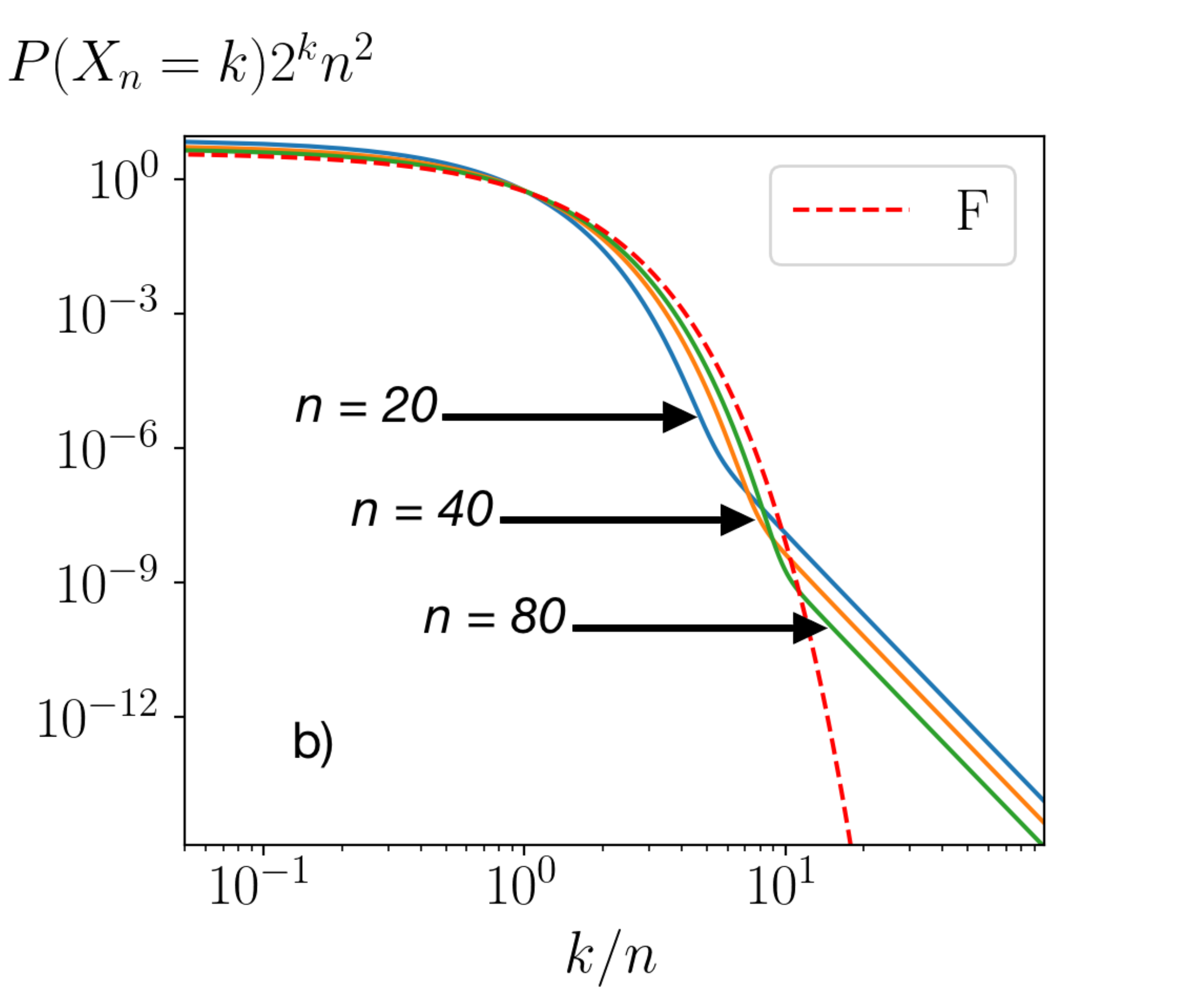}
\ \ \ 
\centerline{\includegraphics[width=8cm]{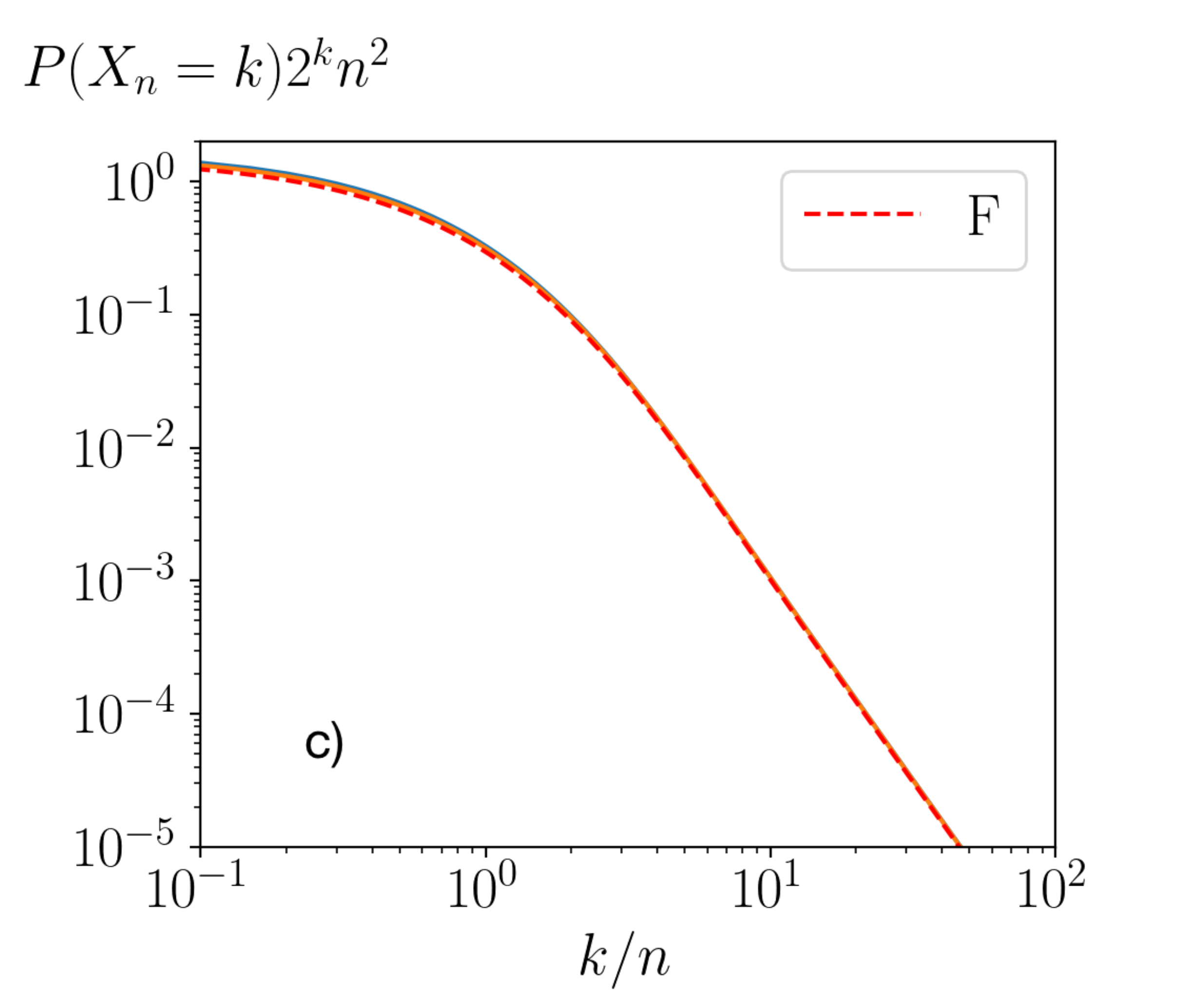}}
\caption{  
    We compare the scaling functions  (\ref{F-scaling}) solutions of (\ref{F-eq}) (dashed lines)   with the solutions of (\ref{recur-dist})  for the same  three cases as in 
 Figure \ref{P(Xn=0)} at times $n=20$, $40$ and $80$. 
 As $n$ increases the convergence to $F$, the  expected scaling function (\ref{expo3}),   is better and better for the two delta peak distribution (figure a) and for the  distribution (\ref{alpha}) with $\alpha=6$ (figure b). For the distribution (\ref{alpha}) with $\alpha=3$ the convergence  is {even faster (figure c) but the scaling function, solution of  (\ref{F-eq}), is different. It is
 the one with $F(0)={3 \over 2}$} (see (\ref{F-large-x})). 
 }
\label{long-time}
\end{figure}

Figure \ref{long-time} compares the  scaling functions solutions (\ref{F-scaling}) solution of (\ref{F-eq}) with the the distributions $P(X_n)$ obtained  by a numerical iteration of (\ref{recur-dist}) with an initial condition which is either  a two delta peaks (\ref{two-delta}) with $p=1/5$ or (\ref{alpha}) for $\alpha=6$ and $\alpha=3$.  In all cases the expected convergence to the scaling function is observed.

\subsection{Sums of exponentials}

One can build another family of exact solutions of (\ref{recursion}) which generalizes (\ref{expo1}): if the initial condition is the sum of  exponentials, then the solution at all times remains a sum of exponentials

\begin{equation}
f(x,t) = \sum_i a_i(t) e^{x \, b_i(t) } 
\label{sum-expo}
\end{equation}
with the parameters $a_i(t)$ and $b_i(t)$ evolving according to

\begin{eqnarray}
\nonumber
\dot{a_i} & = a_i \ b_i +  a_i \sum_{j \ne i} {a_j \over b_i - b_j} \\
\dot{b_i} & = {a_i \over 2}
\label{sum-expo1}
\end{eqnarray}
(The scaling function  (\ref{F-scaling},\ref{particular1}) was an example of such a solution in the case of two exponentials).

It is easy to check that
\begin{equation}
\sum_i a_i - b_i^2 = {\rm  Constant}
\label{inv1}
\end{equation}
 is left invariant by the evolution (\ref{sum-expo1}).
One can also verify that initial conditions on the critical manifold (\ref{manifold1}), i.e. such that
\begin{equation}
\sum_i{a_i \over b_i^2} =1
\label{inv2}
\end{equation}
remain on the critical manifold.
Apart from these two invariants (\ref{inv1},\ref{inv2}), it is in general difficult to integrate the evolution (\ref{sum-expo1}).

Along the critical manifold (\ref{inv2}) however,   one can find the general solution of (\ref{recursion}) when
it  consists of a sum of two exponentials. It is of the form
\begin{equation}
f(x,t) =2  \ {d b_1(t) \over dt} \  e^{x \,  b_1(t) } 
+ 2  \ {d b_2(t) \over dt}  \ e^{x \,  b_2(t) }  
\label{2-expo}
\end{equation}
where $b_1(t)$ and $b_2(t)$  are given by
\begin{equation}
b_{1,2} (t)= -{4 K \over K(t+t_1) \pm \sqrt{ K^2 (t+t_1)^2 - 4 K (t+t_1) \tanh (K(t+t_0)) + 4}}
\label{2-expo-bis}
\end{equation}
where the free parameters $t_0,t_1$ and $K$ could be determined from the initial condition  \ $a_1(0),a_2(0),b_1(0),b_2(0)$.  
(Here  one needs to satisfy the condition (\ref{inv1}) to be on the critical manifold together with $b_1(0)<0$ and $b_2(0)<0$, and this is why  one is left in (\ref{2-expo-bis}) with  only 3  parameters.) 

For real $K, t_0$ and $t_1$ this gives
in the long time limit
$$b_1 \simeq -{2 \over t} \ \ \ ; \ \ \  a_1 \simeq{4 \over t^2} \ \ \ \ \ \ \  {\rm  and} \ \ \ \ \ \ \ b_2 \to -2 K \ \ ; \ \ a_2 \simeq 16 K^2 e^{-2 K (t+t_0)} \ \ .  $$
This shows  that   $a_2$ decays very fast, so that a single exponential  dominates  and the solution becomes given by (\ref{expo3}).

 In fact, on the critical manifold (\ref{manifold1}), 
for any {non negative} intial condition consisting of a finite number of decreasing exponentials,
 one expects that in the long time limit, 
    all the amplitudes $a_i$  decay exponentially except one  and that, asymptotically, the solution is given by (\ref{expo3}). 
\ \\ \ \\ 
{\it Remark:}
by taking $t_1=0$, $t_0=i {\pi \over 2 K} + {8 K^2 \over p } $ and then the limit $K\to 0$ one finds
 $$  b_{1,2} = {3\,  p \,  t^2 \pm \sqrt{288 \, p        t - 3\,  p^2 \, t^4} \over 24-  p  \,      t^3} $$
which corresponds to the initial condition $f(x,0)=       p \, x$.
 One can notice that this solution blows up at a time $t_c=(24/      p)^{1\over 3}$ as expected in (\ref{alpha-less-than-2}) in the case $\alpha=-1$. 
\ \\ \ \\ 
{\it Remark:} the single exponential (\ref{expo1})  is also a particular case of (\ref{2-expo-bis}) as it corresponds to the limit $t_1 \to \infty$.
\ \\ \ \\ 
{\it Remark:} 
in \cite{HMP}, it was also noticed that finite sums of exponentials with time-dependent parameters solve the equation (\ref{HMP1}). Equations similar to (\ref{sum-expo1}) but simpler were also discussed \cite{prosen} in the context of integrable systems and random polynomials.
\ \\ \ \\
\section{Expanding around the scaling functions}
\label{neighborhood}
We have obtained in Section \ref{scaling functions} a one parameter family of solutions (\ref{F-scaling}) which all lie on the critical manifold. In this section we  discuss the evolution of small perturbations around these scaling functions. To do so we consider solutions  of (\ref{recursion}) of the form 
$$f(x,t) = {1 \over (t+t_0)^2} F\left({x \over t + t_0}\right) + \epsilon \  g(x,t)$$
or equivalently 
\begin{equation}
\widetilde{f} (p,t) = {1 \over t+t_0} \widetilde{F} \Big( p (t+t_0) \Big) +  \epsilon \ \widetilde{g} (p , t)  \ \ . 
\label{neighbor}
\end{equation}
with $\widetilde{F}$ given by (\ref{F-Bessel}).
Then at order $\epsilon$,  
 $\widetilde{g}$  evolves according to
\begin{equation}
{d \widetilde{g}(p,t) \over dt} = - g(0,t) + p \, \widetilde{g}(p,t) + {1 \over t+t_0} \widetilde{F} \big( p (t+t_0) \big) \  \widetilde{g}(p,t)  \ \ . 
\label{g-equation}
\end{equation}

 Although this equation is linear, it is non-local in the $q$ variable, because of the presence of $g(0,t)$, and we did not succeed in finding  an explicit solution  for  an arbitrary initial condition $\widetilde{g}(p,0)$.

One can however   obtain  ``eigenfunctions" corresponding to this linear evolution: if one chooses
$g(x,t)$ or $\widetilde{g}(p,t)$
of the form

\begin{equation}
\hspace{-2cm} g(x,t) =(t+t_0)^{\gamma-2} \  G_\gamma\left({x \over t+t_0 } \right)
\ \ \ \ \ \Longleftrightarrow \ \ \ \ \ 
\widetilde{g}(p,t) = {(t+t_0)^{\gamma-1}} \ \widetilde{G}_\gamma\Big(p (t+t_0)\Big) 
\label{G-scaling}
\end{equation}
where $\gamma$ plays the role of an eigenvalue, 
one gets 
 from (\ref{g-equation})  that
$\widetilde{G}_\gamma$ should satisfy
\begin{equation}
q \widetilde{G}_\gamma(q)+ \widetilde{F}(q)\widetilde{G}_\gamma(q)- q \widetilde{G}_\gamma'(q)- (\gamma-1) \widetilde{G}_\gamma(q)= G_\gamma(0)  \ \ . 
\label{G-eq}
\end{equation}
Replacing $\widetilde{F}$ by its expression (\ref{F-Bessel})
one finds for  $\widetilde{G}_\gamma$
\begin{equation}
\label{G-solution}
\widetilde{G}_\gamma(q) = G_\gamma(0)\  q^{-\gamma} \, \left[y\left({q\over 2}\right) \right]^{-2} \int_q^\infty q_1^{\gamma-1} 
\left[y\left({q_1\over 2}\right)\right]^2 d q_1 
\end{equation}
(one has to choose the solution of (\ref{G-eq})  which decays in the limit $q \to \infty$ and this fixes the arbitrary constant in the solution of the linear differential equation (\ref{G-eq})).
For small $q$ one gets formally from (\ref{G-solution}) or directly by solving (\ref{G-eq}) 

\begin{eqnarray}
\widetilde{G}_\gamma(q) = {G_\gamma(0) \over 2 \beta -\gamma}- 
{G_\gamma(0) \ q^2 \over 4 (2 \beta-\gamma ) (2 \beta -\gamma -2) (\beta-1)} + \cdots 
\nonumber
\\ +
c(\beta) \,  q^{2 \beta} \left(  {G_\gamma(0)  \over (\gamma +2) (2 \beta -\gamma )} + \cdots \right)  +
\label{G-small-q}
\\ +  d(\beta,\gamma) \  q^{2 \beta-\gamma}  \left(G_\gamma(0)  + \cdots \right)
\ + \cdots
\nonumber 
\end{eqnarray}
where $c(\beta)$ is given by (\ref{c-beta}) and $d(\beta,\gamma)$  takes different forms depending on the values of $\beta$ and $\gamma$. For example for 
$\gamma- 2 \beta  >0$, one has
$$ d(\beta,\gamma)=  4 \int_0^\infty q_1^{\gamma-1} \left[y\left({q_1\over 2}\right)\right]^2 d q_1 $$
 whereas for  $-1 < \gamma- 2 \beta  <0$, one has
$$ d(\beta,\gamma)=  4 \int_0^\infty q_1^{\gamma-1}\left( \left[y\left({q_1\over 2}\right)\right]^2 -2^{2 \beta -2} \ \Gamma(\beta)^2 \ q_1^{-2 \beta} \right)d q_1 \ \ .  $$
(This is  reminiscent of the various integral expressions of the $\Gamma$ function.)

As in (\ref{F-laplace}) one needs to reorder the terms in (\ref{G-small-q}) by increasing powers of $q$ in the various sectors of $\gamma-2 \beta$ .
\ \\ \ \\ {\it Remark:} for general $\beta$ and $\gamma$ we could not find  expressions of $\widetilde{G}_\gamma(q)$ simpler than (\ref{G-solution}). However, in the case $\gamma=-1$, one can easily check that
\begin{equation}
\widetilde{G}_{-1}(q)
={G_{-1}(0) \over 2 F(0)} 
\Big( \widetilde{F}(q)
-q \widetilde{F}'(q) \Big)
\label{G-1}
\end{equation}
and that in the case $\gamma=0$
\begin{equation*}
\widetilde{G}_{0}(q)= G_0(0) \ {\partial \widetilde{F} (q) \over \partial F(0)}
\ \ . 
\end{equation*}

Also for particular values of  $\beta$ and $\gamma$ 
the eigenfunction $\widetilde{G}_\gamma(q)$ is rational and  one can get some closed expressions. For example
for $\beta$ half-integer and $\gamma=2 \beta-1$
\begin{eqnarray}
\nonumber
\widetilde{G}_2 (q) & = G_2(0) \  { q +4 \over (q+2)^2}  \ \ \ \ \ &  {\rm  for} \ \ \beta={3 \over 2} \\
\label{particular2}
\widetilde{G}_4 (q) & = G_4(0) \  {q^3 + 14 q^2 + 74 q +  144   \over (q^2+6 q +12)^2}  \ \ \ \ \  & {\rm  for} \ \ \beta={5 \over 2} 
\nonumber
\end{eqnarray}
and so on.

A priori, to understand the neighborhood of the scaling solution (\ref{neighbor}), one could try to decompose the initial perturbation $\widetilde{g}(p,0)$ on the eigenfunctions (\ref{G-solution}). We did not succeed in doing this decomposition. One can however analyze the long time behavior of perturbations as $\gamma$ varies.
\\ \ 
\begin{itemize}
\item \underline{For $\gamma <0$,} we see in (\ref{neighbor},\ref{G-scaling}) that the perturbation becomes much smaller than  the dominant term in (\ref{neighbor}). In the particular case $\gamma=-1$ (see 
(\ref{G-1})) the perturbation is nothing but a  shift of order $\epsilon$  of the time $t_0$. So asymptotically the perturbation disappears.
\\ \ 
\item \underline{For $0 < \gamma < 2 \beta-1$} the perturbation is relevant, in the sense that the perturbation in (\ref{neighbor}) grows faster than the dominant term.  Still as long as $\gamma < 2 \beta-1$ the perturbation leaves the solution on the critical manifold (\ref{manifold1}) (because there is no linear term in $q$ in (\ref{G-small-q}) so that (\ref{manifold1}) remains unchanged). It is clear from (\ref{G-small-q}) that the large $x$ decay of the perturbation $G_\gamma(x) \sim x^{\gamma -  2 \beta-1}$ is slower that the decay $F(x)\sim x^{- 2 \beta-1}$ of the dominant  term in (\ref{neighbor}). Therefore one expects   to observe in the long time limit another scaling function $F$, the one corresponding to $2 \beta $ being replaced by $2 \beta-\gamma$. 
\\ \ 
\item \underline{For $\gamma =2 \beta-1$}, the perturbation moves the initial condition away  from the critical manifold and
$$\int_0^\infty x \, f(x,0)  dx =1 + O(\epsilon) \ \ . $$
In this case the perturbation remains small compared to the leading term in (\ref{neighbor}) as long as $\epsilon \  t^\gamma < 1$ . Therefore one expects a critical time $t_c$ to scale with the distance $\epsilon$ to the critical manifold to scale like
$$t_c \sim \epsilon^{-{ 1 \over \gamma}} = \epsilon^{-{1 \over 2 \beta -1}} = \epsilon^{ -{1 \over \alpha-2}} \ \ . 
$$
By repeating  the same argument (\ref{free-energy1})  which led to (\ref{free-energy2})  one can then recover (\ref{critical-behavior-alpha}).
\\ \ 

\item 
 \underline{For $\gamma> 2 \beta-1$}, one has   $f(x,0) \sim g(x,0) \sim x^{\gamma -2 \beta -1 }$  for large $x$ (due to the term $q^{2 \beta-\gamma}$ in  (\ref{G-small-q})) and
one expects a singularity of the form (\ref{alpha-less-than-2}), i.e.
 ${\cal F}_\infty \sim \exp[-\epsilon^{-{1 \over \gamma-2 \beta +1} + o(1)} ] $. 
\end{itemize}

\section{The critical trees}
\label{trees}
To each realization of the process (\ref{recur}) leading to a non-zero value of $X_n$, one can associate a random tree, representing how this value of $X_n$ is obtained. 
This tree connects this value $X_n$ to all  the non-zero  values of $X_k$ which contribute to $X_n$.    It  can be constructed according  to the following recursive rule
for $1 \le m \le n$: one starts at the bottom  of the tree with the value $X_n$. Then  if $X_m$ is a non-zero value on this tree at level $m$,  
\begin{itemize}
\item either there is no branching between level $m$ and $m-1$  and  $X_{m-1} = X_m +1$
with a probability

$$
{\rm Pro}(
X_{m-1}
| X_m)\ = 
 \ {2\,  Q_{m-1} (X_m \, +\ 1) \ Q_{m-1}(0) \over Q_m(X_m)} \ \delta(X_{m-1}, X_m+1)   \ \ . $$
\item or there is a  branching event   between level $m$ and $m-1$   {with probability $
1- 2 \,  Q_{m-1} (X_m+1)  \, Q_{m-1}(0)
/ Q_m(X_m) 
$} leading to two non-zero random values $X_{m-1}^{(1)}$ and $X_{m-1}^{(2)}$ at level $m-1$ with a probability
\begin{equation}
\hspace{-3cm}
{\rm Pro}(
X_{m-1}^{(1)},
X_{m-1}^{(2)} | X_m)= 
{Q_{m-1}( X_{m-1}^{(1)})
 \ Q_{m-1}( X_{m-1}^{(2)}) \over {Q_m(X_m)-2  \, Q_{m-1} (X_m+1) Q \, _{m-1}(0)}}
 \ \delta(X_{m-1}^{(1)} +
X_{m-1}^{(2)} \  ,\  X_m+1)  \ \ .  
\label{split}
\end{equation}
\end{itemize}
It follows immediately from this {construction} that the probability 
 ${\cal P}_{m',m}(X_m)$ that there is no branching up to the level $m'$, if one starts at a value $X_m$ at the $m$-th level of the tree, is

\begin{equation}
 {\cal P}_{m',m}(X_m)  =2^{m-m'}  \ {Q_{m'}(X_m + m-m') \over  Q_m(X_m)} \prod_{\mu=m'}^{m-1} Q_\mu(0) \ \ . 
\label{no-branching}
\end{equation}

Using the scaling form (\ref{sca1}) which relates the discrete problem (\ref{recur}) to the continuous time {equation} (\ref{recursion}) and taking the limit $u \to 0$ one gets from (\ref{no-branching}) that, starting with a value $x$ at time $t$, the probability $\psi_{t',t}(x) $ that there is no branching up to the time $t'$ is
\begin{equation}
{\psi}_{t',t}(x)={f(x + t-t',t')  \over f(x,t)}
\label{no-branching1}
\end{equation}
(note that to leading order in $u$ the product in (\ref{no-branching}) does not contribute).
Similarly, given  that there is a branching at time $t$, the density  probability $\Psi_t(x_1,x_2|x)  $ that a value $x$ splits into two value $x_1$ and $x_2  $ is 
\begin{equation}
 \Psi_t(x_1  ,x_2|x) = {f(x_1   ,t) \    f(x_2,t)  \over  \int_0^x f(x',t) f(x-x',t) dx'} \ \delta(x_1     + x_2     - x)  \ \ .
\label{split-1}
\end{equation}

The  study of the above constructed tree on the critical manifold (\ref{manifold}) was crucial  \cite{6authors}  in  the mathematical proofs of (\ref{critical-behavior})
and of (\ref{critical-behavior-alpha}). On the critical manifold (\ref{manifold}),
 as one expects (see (\ref{sca1},\ref{F-scaling})) that in the large $n$  limit
$$ 2^k \ Q_n(k) \simeq {1 \over n^2} \ F\left({k \over n}\right)$$ with $F$ solution of (\ref{F-eq}),
these expressions become 
\begin{equation}
{\psi}_{t',t}(x)=\left({t \over t'} \right)^2 \ {F({x+t-t' \over t'}) \over F({x\over t})}
\label{no-branching2}
\end{equation}
\begin{equation}
 \Psi_t(x_1  ,x_2|x) = { F({x_1 \over t}) \ F({x_2 \over t}) \over  \int_0^x F({x'\over t}) F({x-x'\over t}) dx'} \ \delta(x_1     + x_2     - x)  \ \ .
\label{split-2}
\end{equation}

For  initial distributions which decay fast enough (i.e. which satisfy (\ref{condition}) for the discrete time problem (\ref{recur}) or (\ref{condition1}) for the continuous time problem (\ref{recursion})) one expects the scaling function $F$ to be given by $F(x)=4 e^{-2 x}$
so that the above expressions (\ref{no-branching2}) and (\ref{split-2}) become
\begin{equation}
{\psi}_{t',t}(x)=\left({t \over t'} \right)^2 \  e^{-{2 (t-t')(x+t)\over t \, t'}} \ \ \ \ \ ; \ \ \ \ 
 \Psi_t(x_1  ,x_2|x) =  {1 \over x} \ \ .
\label{normal}
\end{equation}
This leads to the same critical random trees as those
obtained in
  \cite{HMP} for the problem  (\ref{HMP1}) which can be constructed in the present context as follows:
\begin{enumerate}
\item one starts with a single  particle of mass $\mu_t=x$ at time $t$ at the bottom  of the tree.
\item  then going down in time the mass $\mu_t$ increases linearly $\mu_{t'}=\mu_t+t-t'$ until the first branching event at time $t'$ is reached.

\item   this branching event occurs at rate 
\begin{equation}
{2\mu_{t'} \over  t'^2}
\label{normal1}
\end{equation}  and the mass $\mu_{t'}$ is split uniformly between  two branches. The masses on these two branches   continue to grow and to split independently up to time $0$, in the same way as the branch we started with at time $t$.

\end{enumerate}

\ 

For initial distributions   (on the critical manifold)  which do not satisfy  conditions (\ref{condition}) or (\ref{condition1}), like (\ref{intro-alpha}, \ref{alpha}) for the discrete problem (\ref{recur}) or  $f(x,0) \sim x^{-\alpha}  $ for the continuous problem (\ref{recursion}) {for $2 \le \alpha < 4$}, the above expressions (\ref{no-branching2}) and (\ref{split-2}) remain valid if one chooses the scaling function $F$ solution (see section \ref{scaling functions})  which decays with the same power law  $\alpha$ as the initial condition. 
In this case the critical random trees can also be constructed: 
\begin{enumerate}
\item one starts with a single  particle of mass $\mu_t=x$ at time $t$ at the bottom  of the tree.
\item  then going up  in time the mass $\mu_t$ increases linearly $\mu_{t'}=\mu_t+t-t'$ until the first branching event at time $t'$ is reached
\item   this branching event occurs at rate 
\begin{equation}
- {2 \over t'} - \left( {1 \over t'} + {\mu_{t'} \over  t'^2} \right){F'({\mu_{t'}\over t'} )\over 
F({\mu_{t'} \over t'}) } 
\label{tree2}
\end{equation}
instead of (\ref{normal1}).
The  way the  mass is split remains given by (\ref{split-2}) but is no longer uniform.
\end{enumerate}

In Figure \ref{compar1} we see that the expressions (\ref{no-branching}) obtained by the numerical iteration of (\ref{recur-dist}) converge very well to the expected scaling form (\ref{no-branching2}).

\begin{figure}[h!]
\includegraphics[width=8cm]{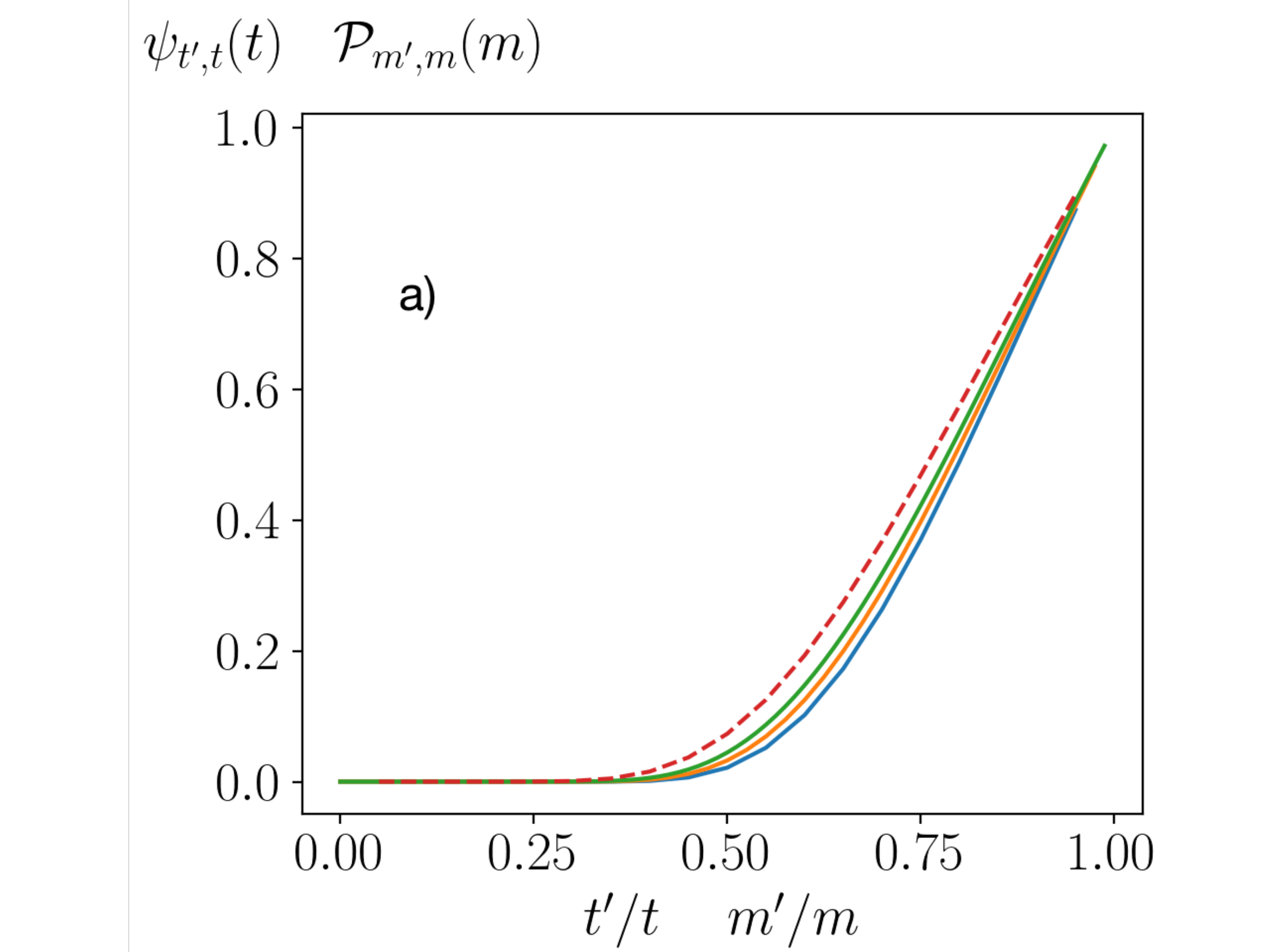}
\ \ \
\includegraphics[width=8cm]{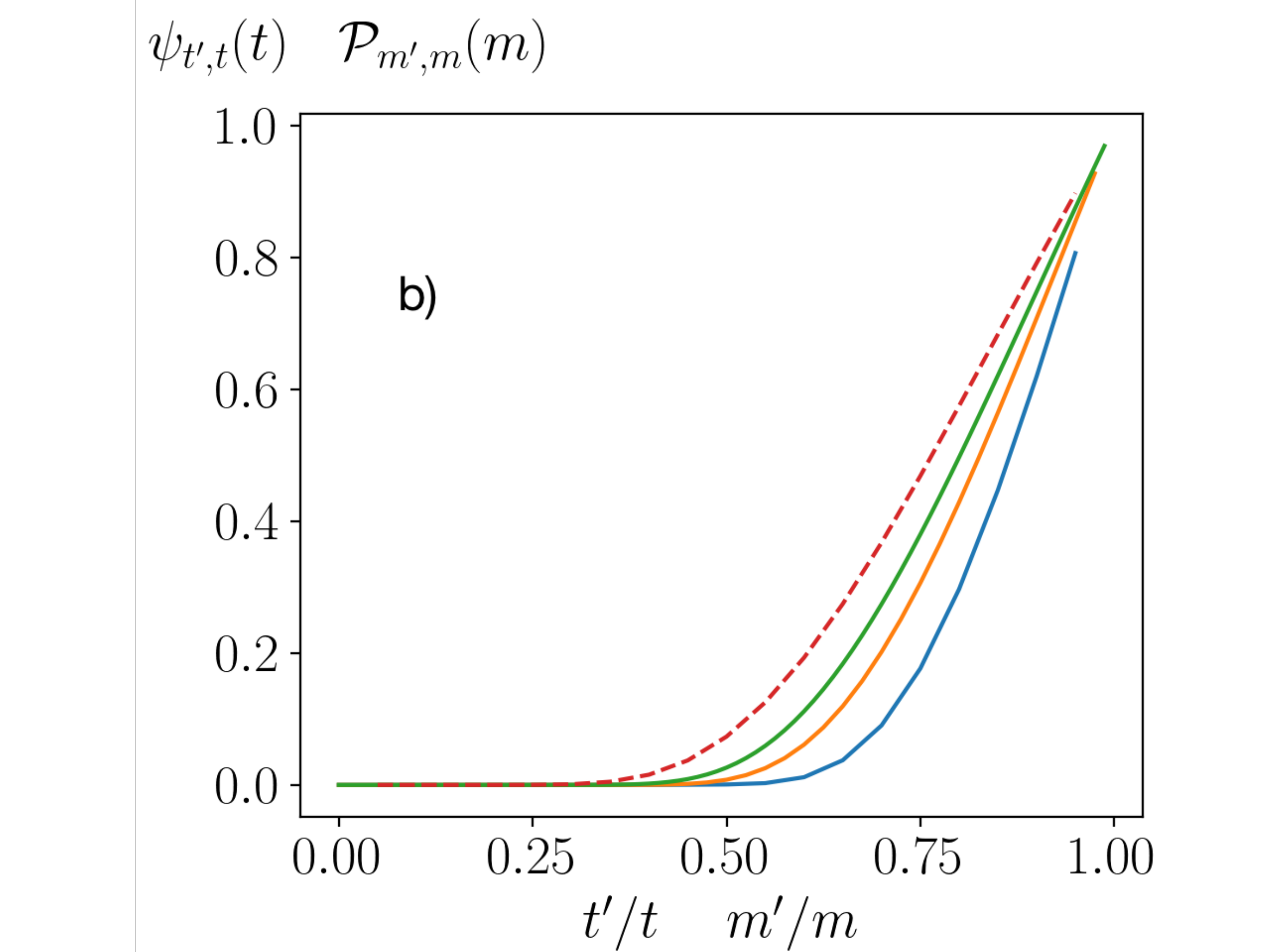}
\ \ \
\centerline{\includegraphics[width=8cm]{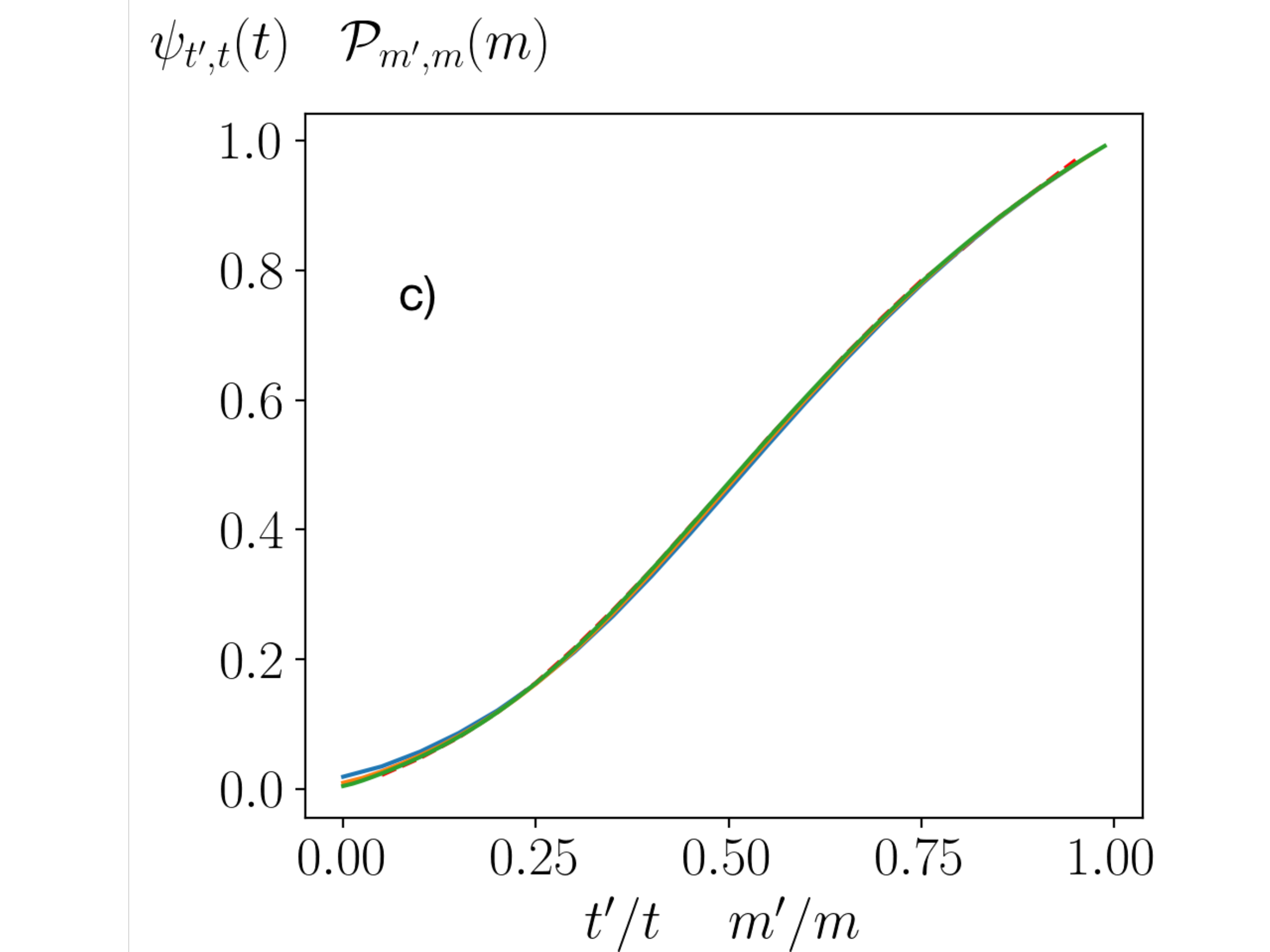}}
\caption{  {
We compare the predictions (\ref{no-branching2}) and (\ref{normal})  with the expressions (\ref{no-branching}) obtained by iterating numerically (\ref{recur-dist}) for the probability of no branching up  to time $m'$  for  two critical  distributions   which satisfy (\ref{condition}) and one which does not
 (the two delta peaks  (\ref{two-delta}) with $p=p_c=.2$  (figure a) and  the distribution  (\ref{alpha})  with  $p=p_c=  1.90956...$  for the case $\alpha=6$ (figure b) and  with $p=p_c=     1.02031..$ for the case $\alpha=3$ (figure c)).
Here we choose as the starting point $X_m=m$.
{In the three cases, we draw the expression (\ref{no-branching}) for $m = 20,40,80$.} The expected convergence is very good if one chooses in each case the appropriate scaling  function  (\ref{normal}) for figures a and b and (\ref{no-branching2}) using   the scaling function $F$ with the right power law decay  (here the one with $F(0)={3 \over 2})$ in the case of figure c.
} }
\label{compar1}
\end{figure}

For the same three distributions as in  Figure \ref{compar1}
one can   also check in Figure \ref{compar2}  the convergence of the splitting probabilities (\ref{split})
to their predicted scaling forms (\ref{split-1}).  

\ 
\begin{figure}[h!]
\includegraphics[width=8cm]{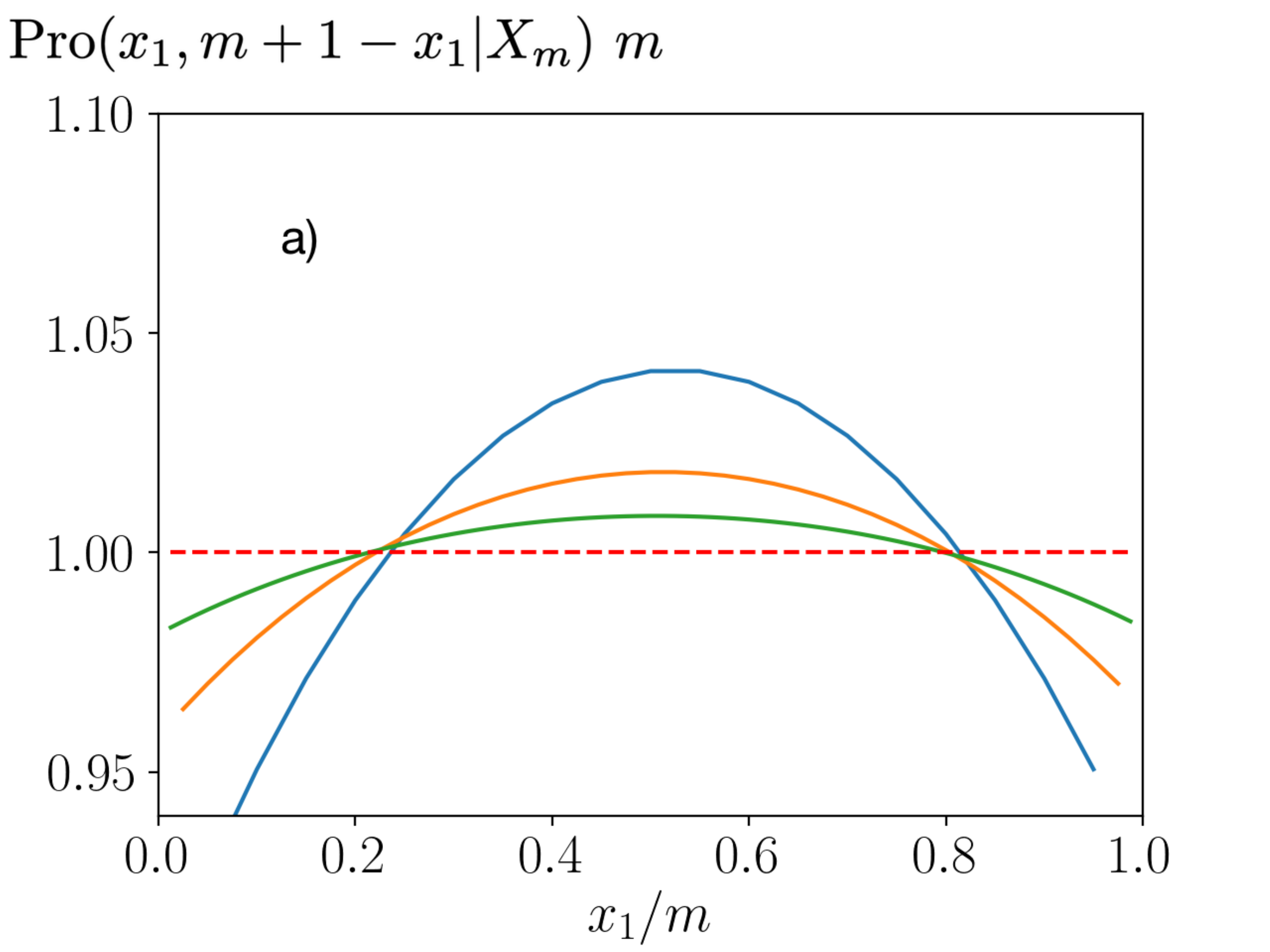}
\ \ \
\includegraphics[width=8cm]{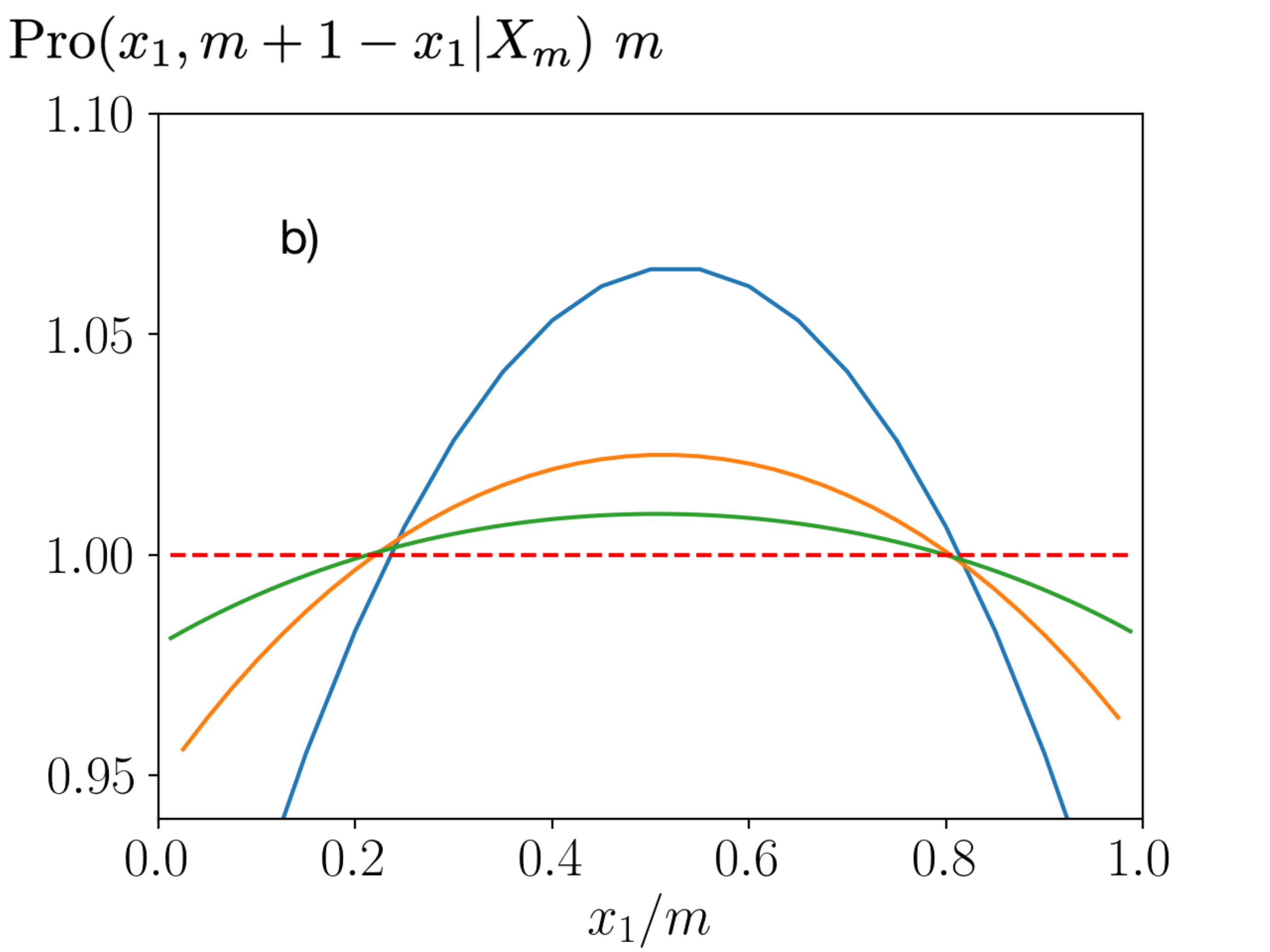}
\ \ \centerline{\includegraphics[width=8cm]{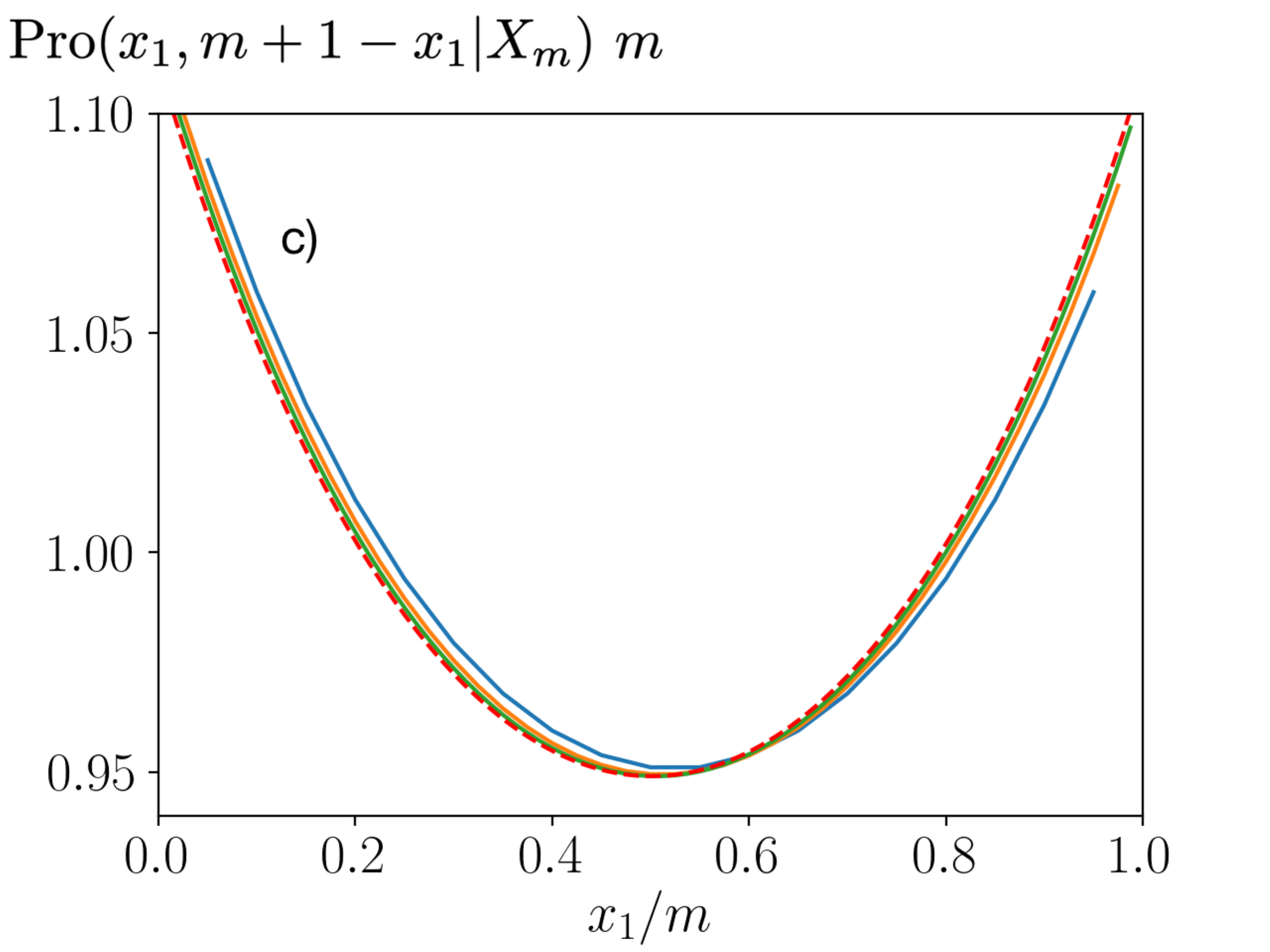}}
\caption{  For the same three distributions as in Figure \ref{compar1} and same sizes ($m=20,40,80$)  we compare the spliting probabilities (\ref{split}) to their scaling forms (\ref{split-2}) which is drawn as a dashed line. We choose the starting point to be $X_m=m$. For the two delta peaks (figure a) and the case $\alpha=6$ (figure b) we observe the expected convergence of $m\ \mathrm{Pro}( x_1,m+1-x_1|X_m)$ to $1$. In the case $\alpha=3$ (figure c) one observes also  a  convergence  but to (\ref{split-2}) with the right choice of $F$, i.e here  : $F(0)=3/2$.
}
\label{compar2}
\end{figure}

\ \\ 
{\it Remark}: all these random  critical trees are scale invariant, in the sense that if all the masses and all the times are multiplied by a fixed factor:
$t \to \lambda t$ and $x \to \lambda x$, they obey the same statistics  as the above constructed trees.
\ \\ \ \\
{\it Remark}:   in the limit $F_0 \to 0$, i.e. $\alpha \to 2$, the expression (\ref{F0-expansion}) gives
$F(x)\simeq F_0/(1+x)^2+  O(F_0^2)$, so that all the branching rates (\ref{tree2}) are of order $F_0$. This means that in this limit, the tree reduces to a single straight line: $\mu_t'= \mu_t + t-t'$. At the next order in $F_0$, one could see also trees with a single branching, and so on: each new order in the small $F_0$ expansion would add trees with an additional branching.

\section{Conclusion}

One of the main progress  of the present work was to
obtain exact expressions  (\ref{F-eq}, \ref{F-Bessel}) for a family of   scaling functions
solutions of (\ref{recursion}).
 These scaling functions   are  indexed by a parameter $\alpha=1+ 2 \beta= 1+ \sqrt{1+ 2 F(0)}$ which controls their power law decay (\ref{F-large-x}).
They generalize the already known scaling function (\ref{expo3})  which was at the origin of the conjecture (\ref{critical-behavior}).  The analysis of the neighborhood of this family of scaling functions in section \ref{neighborhood} confirmed all the other possible critical behaviors  (\ref{critical-behavior-alpha}).
We gave numerical evidence in Figure \ref{F-cur} that these scaling functions are positive on the whole positive real axis when $2 < \alpha \le 4$ and only in this range 
(see one of the  remarks of section \ref{scaling functions}).  We also saw in Figures
 \ref{P(Xn=0)} and  \ref{long-time} that, for the original problem (\ref{recur}), {starting with initial conditions on the critical manifold the distribution properly rescaled}   converges  asymptotically to one of these scaling functions:  initial conditions  whose large $X$  decay is fast enough such as (\ref{two-delta}) or (\ref{alpha})  for $\alpha > 4$ converge to the scaling function (\ref{expo3})
whereas initial conditions with a slower decay converge to the  scaling function $F$ with the same power law decay. Lastly we showed in section \ref{trees}  that to each scaling function one can associate critical trees which generalize the trees found  in \cite{HMP}.

Several aspects discussed in the present paper would need a mathematical proof or further developments, in particular:
\begin{enumerate}
\item   the positivity of the scaling functions for the whole range $2< \alpha <4$. 
\item     the observed  convergence 
(see Figure \ref{long-time})
 of initial conditions on the  critical manifold to the scaling functions.
\item the asymptotics (\ref{factor-alpha})  observed numerically in Figure \ref{P(Xn=0)}.
\item  
what happens in the presence of   extra logarithmic factors 
in the initial distribution
\begin{equation}
P(X_0= k) \simeq {{\rm  Constant} \over 2^k \ k^{\alpha} \ (\log k)^{\alpha'}} \ \ \ \ \ \ \ {\rm  with} \ \  \ 2 < \alpha \le  4 \ \ ?
\label{intro-alpha-1}
\end{equation}
In the particular case  $\alpha=2$ and $\alpha'>1$ where there is still a transition and (\ref{factor-alpha}) 
one expects from the last remark of Section
\ref{trees}  that the tree consists of a single branch which connects $X_n$ to a single leaf $X_0= X_n + n$  implying that
  $(1-P(X_n)) \sim 1/(n^2 \  (\log n)^{\alpha'}) $ instead of (\ref{factor-alpha}).

\item 
it would  be alsointeresting to see what features 
(such as distribution of the number of leaves or  the distribution of coalescence times  \cite{Brunet,Bertoin})  of the 
 critical trees of section \ref{trees} could be computed.
 
\end{enumerate}

The renewed interest for the   problem (\ref{recur}) introduced 35 years ago in the context of 
spin glasses  
\cite{collet-eckmann-glaser-martin2,collet-eckmann-glaser-martin}
originated from attempts to understand the depinning transition in presence of disorder
\cite{giacomin-toninelli,giacomin,alexander,dglt,giacomin-toninelli-lacoin,giacomin_stf,monthus,Berger}
and in particular its version on a hierarchical lattice
\cite{derrida-hakim-vannimenus,tang-chate,giacomin-lacoin-toninelli,derrida-retaux}.
At the end of the present work, it would be interesting to see whether the rich variety of critical behaviors discussed  in \cite{6authors} and here is also present for the depinning problem. In the case of the hierarchical lattice, as explained in \cite{derrida-retaux}, the only difference with the problem (\ref{recur}) is that the $\max$ function  is replaced by a slightly more complicated non-linear function
\begin{equation}
X_n= {\cal G} \left(X_{n-1}^{(1) }+ X_{n-1}^{(2)} \right)
 \ \ \ \ \ {\rm  with} \ \ \ \ 
 {\cal G} (X)= 
X+ \log
\left(\frac{1+(b-1)e^{-X}}{b}\right)  \ \ . 
\end{equation}

One can also  try to generalize (\ref{recursion}) in the following way:
\begin{equation}
{d f(x) \over dt} = {d f(x) \over dx} + {1\over \nu}
\int_0^x \,  f(x_1) dx_1  \cdots
\int_0^x f(x_\nu)  d x_\nu \ \delta(x_1+\cdots + x_\nu-x)  \ \ . 
 \label{recursiona}
\end{equation}
If one looks for scaling functions  as in section \ref{scaling functions} one gets
$$f(x,t) = t^{-{\nu\over \nu-1}} \ F\left({x\over t} \right) $$
where
$F$ satisfies
\begin{equation}
\hspace{-2cm} 
F' 
 +x F' + {\nu \over \nu-1} F
+ {1 \over \nu}
\int_0^x F(x_1) dx_1  \cdots
\int_0^x F(x_\nu)  d x_\nu \ \delta(x_1+\cdots x_\nu-x) =0
\ \ . 
\label{final}
\end{equation}
Trying to determine  for which value of the exponent  $\alpha$ (which controls the large $x$ decay  of the scaling  function $F(x) \sim x^{-\alpha}$)  the solution  of (\ref{final}) is positive, as we did in Figure \ref{F-cur},   we found  numerically  that $1.5 < \alpha  < 2.6$ for the case $\nu=3$.
(For the lower value, 1.5, one can show using the equation satisfied by the Laplace transform that
 it is in general ${\nu \over \nu-1}$ and that $F(0)= {\alpha \over \nu } \left( \alpha - {\nu \over \nu -1} \right)^{1 \over \nu -1}$; on the other hand we  have  no  theory for the value 2.6. We also wonder  how to generalize for $\nu\ge 3$  the equation (\ref{manifold1}) which characterizes the critical manifold).

\appendix
\setcounter{section}{0}

\section{ On the positivity of the solution of $f(x,t)$ solution of (\ref{recursion})}
\label{ap1}

In this appendix we show that if the
initial condition  $f(x,0)\equiv f_{{\rm  initial}}(x)$ is non negative, then the solution $f(x,t)$ of (\ref{recursion}) remains non negative at  any later  time $t$. 
 Let us define the non-decreasing function ${      g}(x)$ by
\begin{equation}
\label{growth1}
{      g}(x) =  \max_{0 \le y \le x} f_{\rm  initial}(y)
\end{equation}
 and a sequence $u_n(x,t)$ of positive functions
by
\begin{eqnarray}
\label{un1}
u_0(x,t)& = f_{{\rm  initial}}(x+t) \\
u_n(x,t)&  = {1 \over 2} \sum_{m=0}^{n-1} \int_0^t d \tau \int_0^{x+t-\tau} dy \ u_m(x+t-\tau-y,\tau) \ u_{n-m-1}(y,\tau)  \ \ .
\nonumber
\end{eqnarray}
Then it is easy to check from (\ref{un1}) that
$$ u_n(x,t) \ge 0$$
and that 
$${d u_n(x,t)  \over dt} = {d u_n(x,t)  \over dx} + {1 \over 2} \sum_{m=0}^{n-1} \int_0^{x} dy \ u_m(x-y,t) \ u_{n-m-1}(y,t)  \  \ . $$
One can also check that
$$0 \le  u_n(x,t) < {t^n \over 2^n}  \    (x + t )^n  \ {      g}(x+t)^{n+1}
$$
This can be shown using the following inequalities 
\begin{eqnarray*}
  \int_0^t d \tau   \, \tau^{n-1} \int_0^{x+t-\tau} dy  \,  {      g}(x + t   -y )^{m+1 }  (x+t-y)^m \ {      g}(y+ \tau)^{n-m} \, (y+ \tau   )^{n-m-1 } &
\\
 \le  {      g}(x+t)^{n+1}\ (x + t    )^{n-1}  \int_0^t\,  d \tau  \ \tau^{n-1} \int_0^{x+t-\tau} \, dy 
\ \ \ \ \ \ \ \ \ \ 
\ \ \ \ \ \ \ \ \ \ 
&
\\  \le  {      g}(x+t)^{n+1} \  (x + t  )^{n}  \int_0^t d \tau  \ \tau^{n-1}
    =
   {      g}(x+t)^{n+1} \  (x + t  )^{n}  
  \   {t^n \over n} \ \ . 
&
\end{eqnarray*}
Therefore at least  when $ t \, (x+t) \, {      g}(x+t)  < 2 $ 
$$f(x,t) = \sum_{n=0}^\infty u_n(x,t)$$
is a convergent series of positive numbers so that   the solution of (\ref{recursion}) is positive and finite.

\section{The unphysical fixed points of (\ref{recursion})}
\label{ap2}
In this appendix we give the expressions of  the fixed points of (\ref{recur}). (The  discussion is simpler although similar to the one on the fixed points of (\ref{rec1}) in \cite{derrida-retaux}). As we will see, none of these fixed points $f(x)$ is positive on the whole positive real axis, so none of them is reachable if the initial condition of (\ref{recur}) is positive.
A fixed point of (\ref{recur}) satisfies
\begin{equation}
 {d f(x) \over dx} + {1\over 2} \int_0^x f(x-y) f(y) d y =0 \ . 
\label{fixed-point}
\end{equation}
 For any  value $f(0)$  one can find a solution  of (\ref{fixed-point}) perturbatively in powers of $x$
$$
 f(x) = f(0) - {f(0)^2 \over 4}  x^2 + {f(0)^3  \over 48} x^4 - {f(0)^4 \over 1152} x^6
+ \cdots  $$
It turns out that  $\widetilde{f}$,  the Laplace transform (\ref{LP}) of  $f$, is solution of 
$$\widetilde{f}^2 + 2 p \widetilde{f} - 2 f(0) =0$$
and one has 
\begin{equation}
 \widetilde{f}(p) = \sqrt{p^2 + 2 f(0)} - p 
\label{LPf}
\end{equation}
which implies that for large $x$
$$f(x) \simeq-  {(8 f(0))^{ 1 \over 4} \over \sqrt{\pi}} {\cos \left(\sqrt{2 f(0)}  \ x  + {\pi \over 4}\right) \over x^{3\over 2} } \ \  .  $$
In fact the fixed point solution can be written as
\begin{equation}
f(x)=    \sqrt{2 f(0)} \ {J_1\big( x \, \sqrt{2 f(0)}\big ) \over x}
\label{Bes1}
\end{equation}
in terms of the Besssel function  $J_1(x)$ 
\begin{equation}
J_1(x) = {1 \over 2 \pi } \int_{0}^{2 \pi} e^{i[t - x   \sin(t)] }  dt \ \ . 
\label{Bes2}
\end{equation}
As this Bessel function    has zeros along the positive real axis,
($x_1=3.832... \ ; \  x_2 \simeq 7.016...$ \ ; \ etc..) and is negative for $x_1 < x < x_2$, none of  the fixed points 
(\ref{Bes1},\ref{Bes2})  can be reached by a solution of (\ref{recursion}) when the initial condition $f(x,0)$ is non negative.

\ \\ \ \\

\end{document}